\RequirePackage{fix-cm}
\documentclass[smallextended,natbib]{svjour3}       
\smartqed  
\usepackage{graphicx}
\usepackage{url}
\let\vec\relax
\DeclareMathAccent{\vec}{\mathord}{letters}{"7E}
\usepackage[colorlinks, allcolors=cyan]{hyperref}
\usepackage{amsmath}
\usepackage{amsfonts}
\usepackage{float}
\usepackage[caption=false]{subfig}
\usepackage{footnote}
\usepackage{algorithm}
\usepackage{enumitem}
\usepackage{xcolor}
\usepackage{multirow,tabularx}
\usepackage{booktabs}
\usepackage{cases}
\newcolumntype{Y}{>{\centering\arraybackslash}X}
\begin{document}

\title{Simulations of the SVOM/ECLAIRs Dynamic Background: A fast, accurate and general approach for wide-field hard X-ray instruments
}

\titlerunning{PIRA : Fast, accurate dynamic background simulations of hard X-ray instrument}       

\author{Sujay Mate \and Laurent Bouchet \and Jean-Luc Atteia \and Arnaud Claret \and Bertrand Cordier \and Nicolas Dagoneau \and Olivier Godet \and Aleksandra Gros \and St\'{e}phane Schanne \and Henri Triou}


\institute{S. Mate \and L. Bouchet \and J-L.Atteia \and O. Godet \at
              IRAP, Universit\'e de Toulouse, CNES, CNRS, UPS, Toulouse, France \\
              \email{Sujay.Mate@irap.omp.eu}\and           
A. Claret \and B.Cordier \and N. Dagoneau \and A. Gros \and S. Schanne \and H. Triou \at
CEA Paris Saclay, IRFU/D\'epartement d'Astrophysique - AIM, 91191 Gif sur Yvette, France
}

\date{Received: date / Accepted: date}

\maketitle

\begin{abstract}
The \textsl{Space Variable Object Monitor (SVOM)} is a forthcoming Chinese - French astrophysics space mission dedicated to the study of Gamma-ray bursts and high-energy transients. ECLAIRs, a wide-field hard X-ray coded mask imager, is the leading instrument for the transient detection and their first localisation. The sensitivity of such instruments is usually limited by the background, either of instrumental or astrophysical origin. Detailed estimations of the background are obtained by simulating the interaction of particles with the matter using, in the present case, the GEANT4 Monte-Carlo toolkit. However, this is a time consuming process, especially when it is needed to carry out all possible geometrical and orbital configurations. Instead, we present a much faster method that allows computing the background in either a static or dynamic (time dependent) way. The method is based on the preliminary calculation of a large particle database using the GEANT4 toolkit followed by a selection process based on the incoming direction and energy of the particles. This approach is as accurate as direct Monte-Carlo methods, while it reduces the computation time by a factor of $10^3 - 10^4$ for our application. We apply this method to compute the SVOM/ECLAIRs dynamic background.

\end{abstract}

\section{Introduction}\label{sec:1}
The \textsl{Space Variable Object Monitor (SVOM)} is a Chinese-French astrophysics space mission~\citep{Cordier2015,Wei2016} with an expected launch in late 2021. The primary focus of the mission is the study of Gamma-ray bursts (GRBs) with prompt detection and multi-wavelength follow-up using space and ground based telescopes. \textit{SVOM} will also respond to gravitational wave events to perform follow-up observations. Onboard \textit{SVOM}, the French instrument ECLAIRs~\citep{Godet2014} will play the key role of autonomously detecting and localising GRBs. This is achieved by using an onboard trigger~\citep{Schanne2014} which will operate in real time and as well a ground trigger operating on the offline data. ECLAIRs is a coded-mask imaging camera with a wide Field-of-View (FoV) of $\sim$2 steradians. It has a pixelated detector plane consisting of 80$\times$80 Schottky-type CdTe semi-conductor detectors with an area of 1024 cm$^{2}$ covering the 4-150 keV energy range~\citep{Lacombe2013,Lacombe2018}. 

The sensitivity of high-energy wide FoV instruments is mainly limited by their high background level. The cosmic X-ray background (CXB)~\citep{Giacconi1962} is the dominant background component for instruments with similar energy range as ECLAIRs~\citep{Churazov2007, Ajello2008, Turler2010}. The CXB photons which are reflected on the Earth's atmosphere~\citep{Churazov2008} (hereafter Reflection) and the Earth albedo (hereafter Albedo) produced by the interaction of cosmic rays~\citep{Sazonov2007,Ajello2008} with the Earth's atmosphere can dominate the background counts when the instrument FoV is completely occulted by the Earth ~\citep{Cordier2008}.

The cosmic rays also contribute to the background. Furthermore, secondary particles (electrons, protons, positrons and neutrons) produced by the cosmic ray interactions with the atmosphere can in-turn interact with the instrument~\citep{Cumani2019}. Moreover, low Earth orbit satellites with orbits such as \textit{SVOM} (altitude of $\sim 623$ km and inclination of $30^\circ$), pass through the high density charged particle (mostly electrons/ protons) region known as the South Atlantic Anomaly (SAA). In this region, the particles fluxes can be as high as $\sim10^4\ \mathrm{electrons \cdot cm^{-2} \cdot s^{-1}}$ and $\sim10^2\ \mathrm{protons\cdot cm^{-2} \cdot s^{-1}}$. Finally, the particles (cosmic rays, SAA particle, etc.) produce many secondaries by interacting with the satellite body. These interactions also make the satellite's elements radioactive (activation phenomenon) causing a delayed emission of photons and/or particles.

The \textit{SVOM} payload will have a nearly anti-solar pointing strategy~\citep{Cordier2008}. This strategy optimises the ground follow up observations of the detected GRBs, but also helps to satisfy few thermal and electric power constraints. Due to this pointing strategy, the Earth transits through the ECLAIRs FoV and produces a modulation of the background count rate. Indeed, an accurate estimation of the background is important to determine the sensitivity of the instrument, to design and develop the onboard and offline triggers and to test the data reduction algorithms. In addition, the presence of the Earth in the FoV has an impact on the image reconstruction. This is an issue for the transient detection and localisation. Hence, it is important to have complete dynamic background simulations taking into account the continuous evolution of the satellite orbital parameters.

Dynamic simulation can be performed using ray-tracing methods, however all the physical mechanisms that lead to the detection of a particle can not be always taken into account. Simulation toolkits like GEANT4 can take all these  physical mechanisms into account, however the simulations of different orbital configurations (e.g position of satellite with respect to Earth, pointing direction, etc.) are time consuming. Therefore, we present a method (named Particle Interactions Recycling Approach or PIRA) with the aim to save computation time by avoiding the necessity to perform new GEANT4 simulations each time the orbital configuration changes.

In Section~\ref{sec:2}, we present the method and the underlying assumptions. In Section~\ref{sec:3}, we present its application to SVOM/ECLAIRs. Finally, in Section~\ref{sec:4}, we summarise the results.


\section{The Particle Interactions Recycling Approach -- PIRA}\label{sec:2}

The PIRA is based on a pre-computed database of particle-instrument interactions simulated using the GEANT4 toolkit (see Section~\ref{sec:2.1.1}). The particles stored in the database are reused to perform new simulations. A selection process based on their incoming direction and energy keeps or discards them for the new simulation. Such an approach retains the accuracy of the GEANT4 simulations, while the recycling of particles bypasses the process of carrying out new simulations each time the characteristics of the simulated particles are changed, in turn saving a large amount of computation time. Hereafter, the word primaries will be used equally for photons and charged particles unless mentioned specifically.

In this section, we discuss the generation of the database and the algorithms used to select the primaries in a static case (fixed  orientation and position of the satellite with respect to Earth) and in a dynamic case (time dependent orientation and position of the satellite). Also, we briefly depict the method of assigning an arrival time to the primaries.

\subsection{Generation of the database}\label{sec:2.1}
The generation of the database is carried out by running GEANT4 simulations by firing large number of input primaries (see Section~\ref{sec:4}). We first outline the GEANT4 simulation setup and then describe the information stored in the database.

\subsubsection{The GEANT4 setup}\label{sec:2.1.1}
GEANT4~\citep{geant4} is a widely used toolkit to simulate particle-matter interactions. It has many applications such as the simulation of instruments for nuclear physics as well as the simulations of X/$\gamma$-ray space instruments (e.g. OSSE/BATSE experiment~\citealt{Shaw2003} or INTEGRAL observatory~\citealt{Ferguson2003}). 

The ECLAIRs pointing axis is set to be the Z-axis of the GEANT4 coordinate frame. The GEANT4 X-Y plane is parallel to the detector plane. Having defined a reference frame,  the GEANT4 toolkit needs two major setups: (a) the instrument mass model (i.e. definition of the geometry, position and chemical composition of each component, see Section~\ref{sec:3.2} for ECLAIRs mass model), and (b) the definition of  the input primary (type of primary, initial position and direction / spatial distribution and energy / spectral distribution).

\subsubsection{The database of primaries}\label{sec:2.1.2}
The database of primaries is generated assuming an isotropic spatial distribution for the input primaries to ease the selection process (see Section~\ref{sec:2.3}). We define a sphere with radius $R_p$ such that the entire instrument is contained within this sphere (hereafter source sphere, see Table~\ref{tab:1} for typical value of $R_p$). The centre of this sphere is placed at the centre and on the top of the detector plane. To increase the efficiency of the Monte-Carlo in terms of computation time, only primaries that have a chance to interact with the instrument are fired. The primaries are generated from the surface of the source sphere towards the instrument within an emission cone having an opening angle $\Gamma_{max}$~\citep[see section 4.2 of][and references therein]{Campana2013}. Integrating the flux over the surface of the entire source sphere ($4 \pi R_p^2$), the rate of simulated primaries is given by,
\begin{equation}\label{eqn:simrate}
 \dot N_s = 4 \pi^2 R_p^2 \Phi \sin^2\Gamma_{max}
\end{equation}
where $\Phi$ (in units of $\mathrm{particles \cdot cm^{-2} \cdot s^{-1} \cdot sr^{-1}}$) is the energy integrated flux of primaries. The extra factor $\pi$ in the overall normalisation comes from the cosine-law biasing~\citep{Campana2013}.

The selection (or rejection) of a primary is based on its incident direction and/or energy. Hence, we first have to store the incident direction ($\theta_p$,  $\phi_p$) and energy of each primary, and secondly, the energy deposition and its location on the detector plane. To reduce the size of the database, we decide to store a primary only if it (or its secondaries) deposits energy in the detector (hereafter detected primary). In the case of a pixelated detector like ECLAIRs, a primary (or its secondaries) can deposit energy either in a single pixel (single event) or in several pixels (multiple events). Hence, we store all the distinct events (positions and energy deposits).

\begin{table*}[ht]
\caption{Example of a source file in the case of ECLAIRs. The coordinates are expressed in the instrument frame. In the case of ECLAIRs, we have $ x_p^2 + y_p^2 + z_p^2 = R^2_p $, where $R_p $ (here= 500  cm) is the source sphere radius (see Section~\ref{sec:2.1.2}).}\label{tab:1}
\begin{tabularx}{\textwidth}{|c|c|*{3}{Y|}*{2}{Y|}}\hline
\multicolumn{1}{|c|}{EventID}&\multicolumn{1}{c|}{Input Energy}&\multicolumn{3}{c|}{Coordinates on the Source Sphere$^\dagger$}&\multicolumn{2}{c|}{Primary Direction}\\
\cline{3-7}
 & (keV) & $x_p$ (cm) & $y_p$ (cm) & $z_p$ (cm) & $\theta_p$ (deg) &$\phi_p$ (deg)\\
\hline
11209 &	10.169 & 269.150 & -33.668 & 420.029 & 34.089 & -8.925 \\[3pt]
1675257 & 27.621 & 62.495 & -191.778 & 457.51 & 25.085 & -72.204 \\[3pt]
... & ... & ... & ... & ... & ... & ... \\
\hline
\end{tabularx}\\
$^\dagger$ These parameters are not mandatory (see Section~\ref{sec:2.3.1.1}).
\end{table*}
\begin{table*}[ht]
\caption{Example of an output event file in the case of ECLAIRs}\label{tab:2}
\begin{tabularx}{.75\textwidth}{|c|c|*{2}{Y|}}\hline
\multicolumn{1}{|c|}{EventID}&\multicolumn{1}{c|}{Deposited Energy}&\multicolumn{2}{c|}{Interaction Position (Pixels)}\\
\cline{3-4}
 & (keV) & X & Y \\
\hline
11209 & 10.169 & 13 & 63\\[3pt]
1675257$^\ast$ & 17.213 & 33 & 66\\[3pt]
1675257$^\ast$ & 10.408 & 33 & 67\\[3pt]
... & ... & ... & ... \\
\hline
\end{tabularx}\\
$^\ast$ {An incident primary may create several secondaries, the latter can deposit energy at different positions (multiple events) in the detector. The energy and position of each of these events are stored.}
\end{table*} 

For convenience, the parameters of the primaries and the energy deposits are stored as separate output files. The parameters related to the primaries are stored in the ``source'' file (Table~\ref{tab:1}). The energy deposits or events, are stored in the ``events'' file (Table~\ref{tab:2}). In both files, the first column is the EventID, a counter identifying the primary, which is used to match the selected primary with its corresponding energy depositions. Furthermore, for each database (i.e each primary type), we define an average efficiency factor as follows:
\begin{align}\label{eqn:eff}
 \bar \eta = \frac{N_{D}^{dat}}{N_{s}^{dat}}
\end{align}
where, $N_{D}^{dat}$ and $N_{s}^{dat}$ are respectively the total number of detected primaries and the total number of input primaries. This factor is used to estimate the number of detected primaries ($N_{D}$) from the number of simulated primaries ($N_{s}$) for all PIRA simulations (see Sections~\ref{sec:2.1.1},~\ref{sec:2.3.1.2} and~\ref{sec:2.3.2}).

For ECLAIRs databases (see Section~\ref{sec:3.3}), the $\bar \eta$ has values lying between {$\sim10^{-3}$} to {$10^{-5}$}. The value depends on the details of the instrument mass model, the type of the primary and its spatial and spectral distribution, and also on the chosen source sphere radius $R_p$.

It is worth mentioning that many others instrumental effects, such as the energy resolution, electronic effects can be applied to the events in post-processing steps.

\subsection{Assignment of an arrival time to each detected event}\label{sec:2.2}
The arrival time of primaries follows the Poisson distribution and the difference between the arrival time of two successive primaries ($dt$) follows an exponential distribution with the probability distribution function:
\begin{equation}\label{eqn:exponpdf}
  f (dt, \lambda ) = 
  \begin{cases}
  \lambda e^{-\lambda  dt}\ & dt \geq 0\\
   0 & dt < 0
  \end{cases}\hspace{15pt}
\end{equation}
where $\lambda$, the rate parameter, usually corresponds to the simulated primary rate $\dot N_s$ (in units of $\mathrm{particles \cdot s^{-1}}$). However, in case of the PIRA, the selection of primaries is based on the detected primaries. Hence, $\lambda$ is given as:
\begin{equation}\label{eqn:simtoint}
 \lambda = \dot N_D = \bar \eta \dot N_s
\end{equation}
where $\bar \eta$ is the average interaction efficiency (Equation~\ref{eqn:eff}) and $\dot N_D$ is the rate of detected primaries. Then the arrival times of the $i^{th}$ and $(i+1)^{th}$ event can be expressed as:
\begin{equation}
 t_{i+1} = t_{i} + dt_{i}
\end{equation}
This relation also holds for a variable rate $\dot N_D(t)$ (see Section~\ref{sec:2.4}). In case of the multiple events (see Section~\ref{sec:2.1.2}), the same arrival time is assigned to all the events produced by the same primary (or its secondaries).

\subsection{Selection algorithms in the static case}\label{sec:2.3}
The core of PIRA are the algorithms used to select the primaries from the database. We distinguish the algorithms based on: (a) selection over the spatial domain  and (b) selection over the energy/spectral domain.

\subsubsection{Selection in spatial domain}\label{sec:2.3.1}
\begin{figure}
\centering
\includegraphics[width=\textwidth]{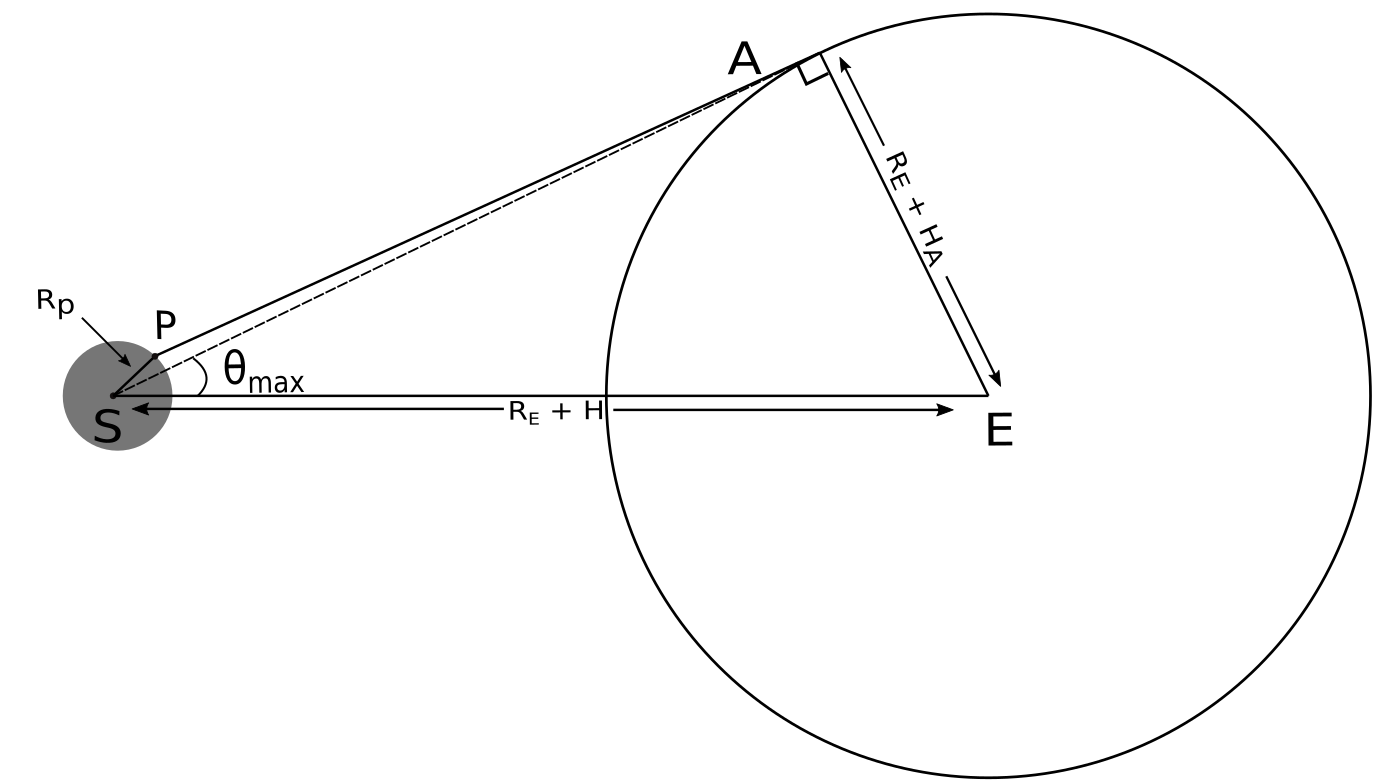}
\caption{Schematic representation of the relative position of the Earth and the satellite. {\bfseries E} is the Earth centre, {\bfseries S} is the satellite position, {\bfseries A} is the point where the trajectory of incoming primary intersects with the atmosphere, $R_E$ is the radius of the Earth, $H_A$ is the height of the atmosphere and $H$ is the satellite altitude (see Section~\ref{sec:3.4} for values of $R_E$, $H_A$ and $H$ in case of ECLAIRs). The grey sphere centred around {\bfseries S} depicts the source sphere with a radius $R_p$. The dashed line represents the case where the size of the source sphere is neglected compared to the satellite altitude i.e. $R_p << H$.}
\label{fig:1}
\end{figure}

We assume the geometrical configuration shown in the Figure~\ref{fig:1}. We define $R_{Eff}$ as the effective Earth radius which is defined as the Earth radius ($R_E$) plus the height of the atmosphere ($H_A$). The satellite altitude is defined as $H$. The point {\bfseries S} is the instrument (satellite) location and the point {\bfseries E} is the centre of the Earth. The point {\bfseries A} is the intersection of the incoming primary's trajectory with the atmosphere. All the coordinates are defined with respect to the instrument frame (see Section~\ref{sec:2.1.1}).  

There are two ways to select primaries in the spatial domain. The first relies only on the incident  direction of the primaries (hereafter directional selection, see Section~\ref{sec:2.3.1.1}) and the second one relies on a directional selection combined with a possible modification of original spatial distribution (hereafter spatial mapping, see Section~\ref{sec:2.3.1.2}). Here, we assume that the database has been generated using a specific energy distribution of the input primaries and it is unmodified by the selection process.

\paragraph{Directional selection\\}\label{sec:2.3.1.1}
This selection process is intended to be applied to the CXB photons (see Section~\ref{sec:3} for an application to ECLAIRs) and the cosmic-ray particles. To use this selection process for cosmic-rays, we have to neglect the East-West anisotropy effect. We assume that the distribution of primaries is isotropic on the sky and that the presence of the Earth shields the primaries coming from that direction. Hence, the estimation of the background is done by a selecting the primaries according to their incoming directions (reject all the primaries which intercept the Earth). The rate at which primaries should be simulated ($\dot N_s$) from the source sphere is given by equation~\ref{eqn:simrate}. Since, only the information about the detected primaries is stored in the database, this rate is translated into the rate of detected primaries $\dot N_D$ as defined in Equation~\ref{eqn:simtoint}.

For each primary crossing the Earth, we have the following relation (Figure~\ref{fig:1}):
\begin{equation}
    \vec{SA} + \vec{AE} = \vec{SE} 
\end{equation}
where,
\begin{align}
 \vec{SA} =&\ \vec{SP} + \vec{PA}
          =\begin{pmatrix}
                x_p\\
                y_p\\
                z_p
             \end{pmatrix}
        + 
        \beta \begin{pmatrix}
                u_{x}\\
                u_{y}\\
                u_{z}
             \end{pmatrix}\\[5pt]
 \vec{SE} =& \begin{pmatrix}
                x_E\\
                y_E\\
                y_E
             \end{pmatrix}
           = (R_E + H) \begin{pmatrix}
                \sin\theta_E\cos\phi_E \\
                \sin\theta_E\sin\phi_E \\
                \cos\theta_E
             \end{pmatrix}\\[5pt]
\intertext{with,}
& \begin{pmatrix}
    u_{x}\\
    u_{y}\\
    u_{z}
  \end{pmatrix} =  
  \begin{pmatrix}
     \sin\theta_p\cos\phi_p \\
     \sin\theta_p\sin\phi_p \\
     \cos\theta_p
  \end{pmatrix}\\[5pt]
\intertext{and,}
&\|\vec{AE}\| = R_{Eff} = R_{E} + H_{A}
\end{align}
We obtain the following quadratic equation in $\beta$:
\begin{equation}\label{eqn:dirn}
    (x_p + \beta u_{x} - x_{E})^2 + (y_p + \beta u_{y} - y_{E})^2 + (z_p + \beta u_{z} - z_{E})^2 = R_{Eff}^2\\[5pt]
 \end{equation}
where $\theta_E$ and $\phi_E$ are the zenith and azimuth angles corresponding to the Earth centre direction; $\theta_p$ and $\phi_p$ are the incident primary directions; $x_p,\ y_p,\ z_p$ are the primary coordinates on the source sphere (Table~\ref{tab:1}). The parameter $\beta$ is the length of the vector $\vec{AP}$. If equation~\ref{eqn:dirn} has a positive solution in $\beta$, then the trajectory intersects with the Earth and consequently the primary is rejected. For ECLAIRs (see Section~\ref{sec:3}), we use this method to select the photons of the CXB background from the database.

The  incident primary position is not mandatory and hence not necessarily stored in database source file (see Table~\ref{tab:1}). In such a case, another way of selecting primaries is used. It relies on the assumption that the Earth radius is negligible compared to the satellite altitude ($R_p << H $, see Figure~\ref{fig:1}). It involves a coordinate transform of all the incoming primary directions ($\theta_p$, $\phi_p$). This is achieved by applying a rotation to the primary direction so that the attached $z$ - axis of the frame is aligned with the direction $\vec{SE}$. In the rotated frame, primaries which have zenith angle less than $\theta_{max}$ (defined by the cone produced by the directions $\vec{SE}$ and $\vec{SA}$) are discarded as their direction intercepts the Earth. This alternative method produces results comparable to those obtained using Equation~\ref{eqn:dirn} with discrepancy of only a few primaries over a million.

For duration $\mathrm{\Delta} t$, we pick $N_D = \dot N_D \mathrm{\Delta} t $ primaries from the database. We assign an arrival time to each selected primary according to the rate parameter $\dot N_D$ (Equation~\ref{eqn:simtoint}). The final number of detected primaries is obtained by selecting or rejecting each detected primary following the steps given in algorithm~\ref{algo:1}. For the first step of the algorithm, a primary is picked up randomly from the database and then discarded from the list of available database primaries. In doing so, a primary is only picked once during the simulation. A simple method to achieve this is to perform a random permutation of the primaries in the database and select them sequentially. This way of picking primaries is also used for the algorithms discussed in the following subsections. 

This method of selecting primaries based on the direction is essentially the same as the post-processing step used by~\cite{Zhao2012} to estimate the CXB background component of ECLAIRs when the Earth is in the instrument FoV.

\begin{algorithm}
  \caption{Directional selection algorithm}\label{algo:1}
  Start with $t=0$ and while $t < \mathrm{\Delta} t$ do,
  \begin{enumerate}[leftmargin=1cm]\parskip5pt
    \item Pick up a primary randomly from the database source file and discard it from the list of available database primaries. The arrival time of the primary is set to $t$.
    \item Find whether the primary direction intercepts the Earth based on the directional information (Equation~\ref{eqn:dirn}).
    \item If the primary crosses the Earth reject it and go to step 5.
    \item If not, then match the primary EventID in the events file and record the detected event(s)
    \item Compute $dt$ (Equation~\ref{eqn:exponpdf}) and the arrival time of the next primary as: $t = t + dt$ 
  \end{enumerate}
\end{algorithm}

\paragraph{Directional selection with spatial distribution mapping\\}\label{sec:2.3.1.2}
This process transforms an initial spatial distribution of primaries into another distribution. Here, this spatial distribution mapping is applied to the Earth Albedo emission.

\begin{figure}
\centering
\includegraphics[width=.9\textwidth]{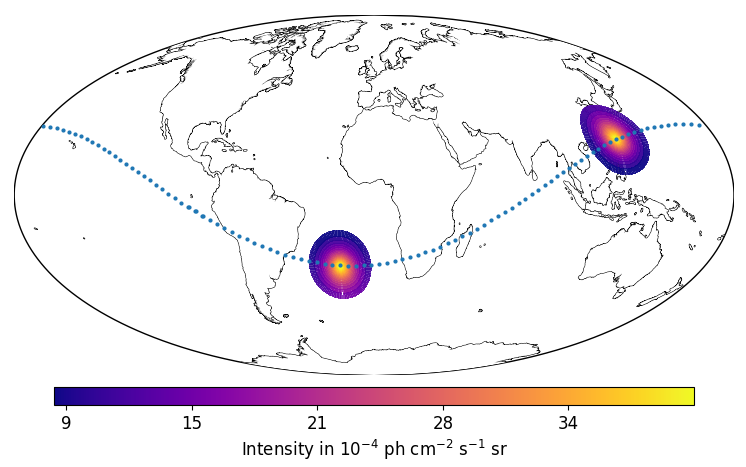}
\caption{The Albedo intensity (as seen from the satellite position) at two points during a typical \textit{SVOM} orbit. The blue dots represent the path of \textit{SVOM}. The circular region shows the part of the atmosphere seen by the satellite at the given altitude ($\sim600$ km) and given position. Within this region the intensity has a directional dependence (Equation~\ref{eqn:alb2}).}
\label{fig:alb_var_orb}
\end{figure}

\cite{Sazonov2007} have derived a convenient formula to model this emission which is valid for satellite altitudes above $\sim100$ km. The Albedo photons are emitted from the Earth's atmosphere with a direction dependent intensity. It also depends on the satellite position in latitude and longitude (Section~\ref{sec:3.1.3}). For \textit{SVOM}, an atmospheric cap of radius $\sim$ 2400 km is visible from the satellite position. The variation of the Albedo intensity within this part of the atmosphere is shown in Figure~\ref{fig:alb_var_orb} for two different positions on a typical orbit. 

From the satellite's point of view, the source sphere surface has a concave shape while the Earth's atmosphere has a convex shape. To take this into account, we need to relate the distributions of photons on the atmosphere cap and on the source sphere.

\begin{figure}
\centering
\includegraphics[width=.9\textwidth]{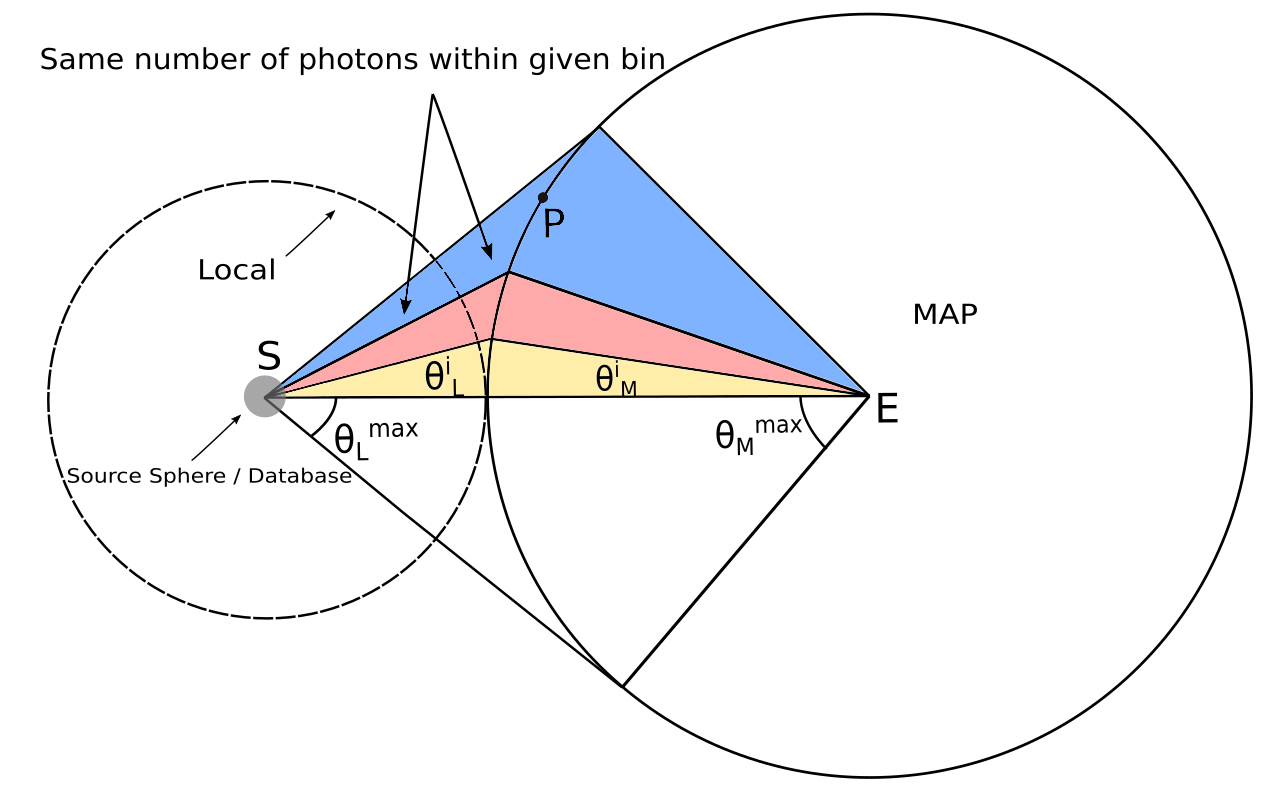}
\caption{Schematic diagram showing the geometrical configuration used to pixelate the visible part of the Earth's atmosphere to calculate the Albedo emission. {\bfseries S} and {\bfseries E} have the same meaning as in Figure~\ref{fig:1}. The ``Local'' sphere (dashed line) is the expanded version of the ``Database" source sphere. The source sphere (database) is shown as the grey circular region. In the ``Local'' frame, the visible part of atmosphere (as seen from satellite) subtends a spherical cap of half angle $\theta_L^{max}$. The corresponding half opening angle in ``MAP" frame is $\theta_M^{max}$. $\theta_L^i$ and $\theta_M^i$ are the boundaries of the angular bin $i$ in ``Local" and ``MAP" frames respectively. To illustrate the assumption made in Section~\ref{sec:2.3.1.2}), same colours are used to show that the number of photons emitted from the Earth Atmosphere towards  the direction of the satellite is the same in the ``Local" and ``MAP" frames.}
\label{fig:albmap}
\end{figure}

To begin, we define three surfaces (Figure~\ref{fig:albmap}), the first is the GEANT4 source sphere (or ``Database"), the second is the ``Local" surface and the third is the ``MAP" surface which is the Earth's atmosphere. For our purpose, the ``Local" surface is an expanded version of the source sphere surface. Then, we define two coordinate frames, the first one is the ``Local'' frame centred at the satellite position {\bfseries S} with the $Z_L$-axis oriented toward the direction of the  Earth centre (direction $\vec{SE}$). The second one is the ``MAP'' centred at the Earth's centre {\bfseries E} with the $Z_M$-axis oriented towards the satellite (direction $\vec{ES}$, Figure~\ref{fig:albmap}). In both frames, the X-Y plane is perpendicular to the corresponding Z-axis.

We assume that the satellite dimensions being small with respect to the satellite altitude ($R_p << H$), it can be approximated as a point source (see Figure~\ref{fig:albmap}). The visible part of atmosphere, as seen from the satellite, subtends a spherical cap of half opening angle $\theta_L^{max}$ in the ``Local" frame. The same cap subtends a half opening angle $\theta_M^{max}$ in the ``MAP" frame. We define the zenith $\theta_L$ ($\theta_M$) and azimuth $\phi_L$ ($\phi_M$) angles in  the ``Local'' (``MAP") frames. The coordinates of any point $P$ (Figure~\ref{fig:albmap}) on the atmosphere spherical cap in the ``MAP" frame (Figure~\ref{fig:albmap}) and in the ``Local" frame are related by:

\begin{equation}
    \cos \theta_{M} = \frac{\left(1-l \times \cos \theta_{L}\right)}{k}\\
\end{equation}
where,
\begin{align}
    l &= \cos \theta_{L}-\sqrt{k^{2}-\sin^2 \theta_{L}}\\
    k &= \frac{R_E + H_A}{R_E + H}
\end{align}
and,
\begin{equation}
        \phi_M = \phi_L - \pi + \psi
\end{equation}
where $\psi$ is a roll angle and for convenience we take $\psi = 0$. To convert the emission intensity from the ``MAP" frame  to the ``Local" frame, we discretize the spherical caps into $\mathcal{N}_{\theta} \times \mathcal{N} _{\phi}$ angular bin. Each angular bin ($i, j$) is then delimited, in the ``Local" frame, by the zenith angles [$\theta_L^{i-1}$, $\theta_L^{i}$] and azimuth angles [$\phi^{j-1}_L$, $\phi^{j}_L$] and in the ``MAP'' frame by [$\theta_M^{i-1}$, $\theta_M^{i}$] and [$\phi^{j-1}_M$, $\phi^{j}_M$]. 

Secondly, we assume that the photons emitted towards the direction of the satellite from a given angular bin ($i,j$) in ``MAP" frame are the same in the corresponding angular bin ($i,j$) in the ``Local" frame. This is shown in Figure~\ref{fig:albmap} where the flux in each of these bins is represented with the same colour. The effective cross-section of the source sphere as seen from ``MAP" frame is given by $\mathrm{\pi R_p^2}$. Assuming the source sphere is infinitesimally small, we can use the relation given in~\cite{Juul1976}~\citep[see also][]{Watts1965} to obtain the photon rate $\dot N_{ij}^{map}$ (in units of $\mathrm{ph \cdot s^{-1} }$) in each angular bin as:
\begin{equation}\label{eqn:map}
 \dot N_{ij}^{map} = \pi R^2_p k^2 \int^{\theta^{i+1}_M}_{\theta^{i}_M}\int^{\phi^{j+1}_M}_{\phi^{j}_M} I(\theta,\,\phi)\frac{\cos \theta-k}{\left(1+k^2-2k\cos \theta\right)^{3 / 2}}\sin\theta\ d\theta\ d\phi
\end{equation}
Here $I(\theta,\,\phi)$ is the Albedo emission intensity (in units of $\mathrm{ph \cdot cm^{-2} \cdot s^{-1} \cdot sr^{-1}}$). Furthermore, we assume that the spatial and energy distributions of the photons on the ``Local'' sphere (sphere with radius $R_L$) and on the source sphere (``Database") are identical. By definition, ``Local" surface is an extended version of the source sphere (with a photon flux ratio of $\frac{R^2_p}{R^2_L}$) and hence the angular bin size is the same in both of these frames. The photon rate (in units of $\mathrm{ph \cdot s^{-1}}$) in each angular bin in ``Local" frame is given by:
\begin{align}
    \dot N_{s(ij)} = \frac{\dot N_s} {4\pi} \times \int^{\theta^{i+1}_L}_{\theta^{i}_L}\int^{\phi^{j+1}_L}_{\phi^{j}_L} \sin\theta\ d\theta\ d\phi
\end{align}
where $\dot N_s$ is the photon rate on the entire source sphere (rate of primaries). We choose $\dot N_s$ such that the photon rate in the ``Local" and ``MAP" frame satisfy following condition for each angular bin:
\begin{equation}\label{eqn:ratio}
  r_{ij} = \frac{\dot N_{ij}^{map}}{\dot N_{s(ij)}} \leq 1
\end{equation}
This condition ensures that the number of photons per angular bin in the ``Local" frame is always larger than the number of photons per angular bin coming from the ``MAP" frame. To have the same number of photons per angular bin in both frames, we keep only a fraction $r_{ij}$ of the database photons. For convenience, we take equal area angular bins on the ``Local" cap; this ensures the same photon rate in each angular bin. The angular size of each bin is defined as:
\begin{equation}
  A_{ij} =  A = \frac{2 \pi (1 - \cos \theta_L^{max})}{\mathcal{N}_{\theta}\mathcal{N}_{\phi}}
\end{equation}
Then the photon rate is  given as:
\begin{equation}
\dot N_s = \frac{2\ \mathcal{N}_{\theta} \mathcal{N}_{\phi}}{(1 - \cos \theta_L^{max})} \times max( \dot N_{ij}^{map}) 
\end{equation}
The rate of detected photons $\dot N_D$ is calculated using the average efficiency given by Equation~\ref{eqn:eff}. For a duration $\mathrm{\Delta} t$, we need to pick $N_D = \dot N_D \mathrm{\Delta} t $ photons from the database. We assign an arrival time to each photon according to the rate parameter $\dot N_D$ (Equation~\ref{eqn:simtoint}). The photons picked from the database are selected by applying a directional selection and a decimation process (in order to adjust the number of photons) as described in Algorithm~\ref{algo:2}.

In order to select photons from the database, each photon direction ($\theta_p$, $\phi_p$) is expressed into the ``Local" frame coordinate system (step 4 of Algorithm~\ref{algo:2}). This transformation is performed using the inverse of the rotation matrix corresponding to the ZYZ rotation rule~\citep{Henderson1977} and defined by Euler angles ($\phi_E$, $\theta_E$, $\psi_E$). The angle $\psi_E$ is the roll angle (see Section~\ref{sec:3.5}) which depends on the detector plane orientation with respect to the North celestial direction. This rotation is similar to the alternative method to select primaries based on their incoming direction described in  Section~\ref{sec:2.3.1.1}.

\begin{algorithm}
  \caption{Directional selection with spatial distribution mapping}\label{algo:2}
  Start with $t=0$ and while $t<\mathrm{\Delta} t$ do,
  \begin{enumerate}[leftmargin=1cm]\parskip5pt
    \item Pick up a photon randomly from the database source file and discard it from the list of available photons in the database. The arrival time of the photon is set to $t$.
    \item Find whether the photon direction intercepts the Earth (Equation\ref{eqn:dirn}). 
    \item If the photon crosses the Earth, select it. Otherwise go to step 9.
    \item If the photon is selected, then find the angular bin ($i$, $j$) on the ``Database / Local'' surface to which the photon corresponds.
    \item Compute ratio ($r_{ij}$) using Equation~\ref{eqn:ratio}.
    \item Draw a random number ($u$) from an uniform distribution.
    \item If $r_{ij} > u$, then select the photon. Otherwise reject it and go to step 9.
    \item If the photon is selected then match the photon EventID in the events file and record the detected event(s) with time $t$.
    \item Compute $dt$ (Equation~\ref{eqn:exponpdf}) and the arrival time of the next photon as: $t = t + dt$.
  \end{enumerate}
\end{algorithm}

It should be noted that, in the above formula (Equation~\ref{eqn:map}), the source sphere is assumed to be infinitesimally small. To take into account the source sphere dimension, the exact formula for the ``view factor" between two spheres given in by~\cite{howell2015thermal}~\footnote{see online catalog at \url{http://www.thermalradiation.net/sectionb/B-67.html}} can be used.

Estimation of the Reflection background~\citep[as approximated by][]{Churazov2008}  can be considered as a peculiar case where the emission intensity $I(\theta,\,\phi)$ is constant.

\subsubsection{Selection in the spectral domain}\label{sec:2.3.2}
This selection process is applied to the charged particles of the SAA. The flux and the spectral distribution of these particles vary as the satellite passes through the SAA. The SAA spans a large region from about $50^\circ$S to $0^\circ$ in latitude, $90^\circ$W to $40^\circ$E in longitude with an altitude extent from $\sim200$ km to $\sim 6000$ km. Hence, the presence of the Earth in the FoV does not affect this local particle distribution. To further simplify the modelling, the pitch angle of particles is ignored, the spatial distribution of the particles is assumed to be isotropic around the satellite. The selection of primaries in this case is equivalent of modifying the energy distribution (shape and intensity) of the database primaries. 

We define the quantity $\mathbf{s}(E)^{dat}$, which is the total database spectrum $S(E)^{dat}$ (in units of $\mathrm{particles \cdot cm^{-2} \cdot keV^{-1}}$) normalised to one incident primary as:
\begin{equation}
  \mathbf{s}(E)^{dat} = \frac{S(E)^{dat}}{N_\circ}\hspace{10pt} \ [\mathrm{cm^{-2} \cdot keV^{-1}}]\\[5pt]
\end{equation}
Here $E$ is the energy and $N_\circ$ is the total number of primaries of a given type drawn to generate the database. The spectrum, that we want to simulate, $S(E)$ can be rewritten in term of a re-sampled version of the database spectrum as:
\begin{equation}\label{eqn:ratioCP}
     S(E) = r(E) \dot N_s  \mathbf{s}(E)^{dat}\\[5pt]
\end{equation}
where $r(E)$ is an energy dependent factor and $\dot N_s$ (in units of $\mathrm{particles \cdot s^{-1}}$) is the re-sampling rate of the database primaries. At each energy, we need to pick at least the same (or a larger) number of primaries from the database than required for the simulation. This implies that $r(E) \leq 1$ at each energy. Then the minimum required database re-sampling rate is given as:
\begin{equation}
    \dot N_s =  max\left(\frac{S(E)}{\mathbf{s}(E)^{dat}}\right)
\end{equation}
We express the ratio given in Equation~\ref{eqn:ratioCP} in terms of the detected primary rate $\dot N_D$ (see Section~\ref{sec:2.1.2}). To do so, for a given energy $E$, we define the primary interaction efficiency ($\eta(E)$) as the ratio of the number of detected primaries ($N(E)_D$) to the number of drawn primaries ($N(E)_s$).
\begin{equation}
 \eta (E) = \frac{N(E)_{D}}{N(E)_{s}}
\end{equation}
This ratio does not depend on the assumed spectral shape. Then $r(E)$ from Equation~\ref{eqn:ratioCP} can be rewritten as:
\begin{equation}
    \dot N(E)_D = r(E) \dot{\mathcal{N}}^{dat}_{D}
\end{equation}
where $\dot N(E)_D$ is the rate of detected primaries at energy $E$ for the spectrum $S(E)$ and $\dot{\mathcal{N}}^{dat}_{D}$ is the rate of detected primaries at energy $E$ for the re-sampled database spectrum. Then, the total rate of detected primaries $\dot N_D$ can be written as:
\begin{equation}
\dot N_D =  \int_{0}^{\infty} \dot{\mathcal{N}}^{dat}_{D}\,\mathrm{d}E =\bar \eta \dot N_s 
\end{equation}
where $\bar \eta$ is the average primary interaction efficiency defined in Equation~\ref{eqn:eff}. For a duration $\mathrm{\Delta} t$, we pick $N_D = \dot N_D \mathrm{\Delta} t $ primaries from the database. We assign an arrival time to each primary according to the rate parameter $\dot N_D$ (Equation~\ref{eqn:simtoint}) and select (accept or reject) these primaries as described in Algorithm~\ref{algo:3}.
\begin{algorithm}
  \caption{Selection in the Spectral Domain}\label{algo:3}
  Start with $t=0$ and while $t<\mathrm{\Delta} t$ do,
  \begin{enumerate}[leftmargin=1cm]\parskip5pt
    \item Pick up a primary randomly from the database source file and discard it from the list of available primaries. The arrival time of the primary is set to $t$.
    \item Compute ratio $r(E)$ using Equation~\ref{eqn:ratioCP}.
    \item Draw a random number ($u$) from an uniform distribution.
    \item If $r(E) > u$ then select the primary. Otherwise reject it and go to step 6.
    \item Match the primary EventID in the events file and record the detected event(s) along with arrival time $t$.
    \item Compute $dt$ (Equation~\ref{eqn:exponpdf}) and the arrival time of the next primary as: $t = t + dt$.
  \end{enumerate}
\end{algorithm}

\subsection{Modification of the algorithms to perform dynamic simulations}\label{sec:2.4}
In this case, the orbital parameters (Earth directions, satellite altitude etc.) depend on time. Hence, the parameters of the algorithms depend on the primary arrival time, which is incremented each time a primary is extracted from the database. Hence, at each time $t$, we compute the detected primary rate ($N_D(t)$) and if required the ratio, either $r_{ij}(t)$ (Equation~\ref{eqn:ratio}) or $r(E,t)$ (Equation~\ref{eqn:ratioCP}). The remaining steps to select the primaries are the same as those given in Algorithms~\ref{algo:1},~\ref{algo:2} and~\ref{algo:3}

\subsection{Background due to the activation of the satellite body}\label{sec:2.5}
We do not consider the activation background component. Its accurate estimation heavily depends on the impurities present in the material which are generally poorly known at the pre-launch stage~\cite[preliminary analysis for ECLAIRs done by][]{Bringer2013}. During the flight, this component contains various activation lines, which can be identified in the measured background spectrum~\cite[see][for INTEGRAL/SPI and Astro-H respectively]{Weidenspointner2003, Odaka2018}.

\section{Application to SVOM/ECLAIRs}\label{sec:3}
We perform simulations to compute the CXB, Reflection, Albedo and SAA background in both static and dynamic cases. We first describe the models used to compute the emission of these background components.

\subsection{Emission models for the background components}\label{sec:3.1}
\begin{figure}[h]
\centering
\includegraphics[width=0.8\textwidth]{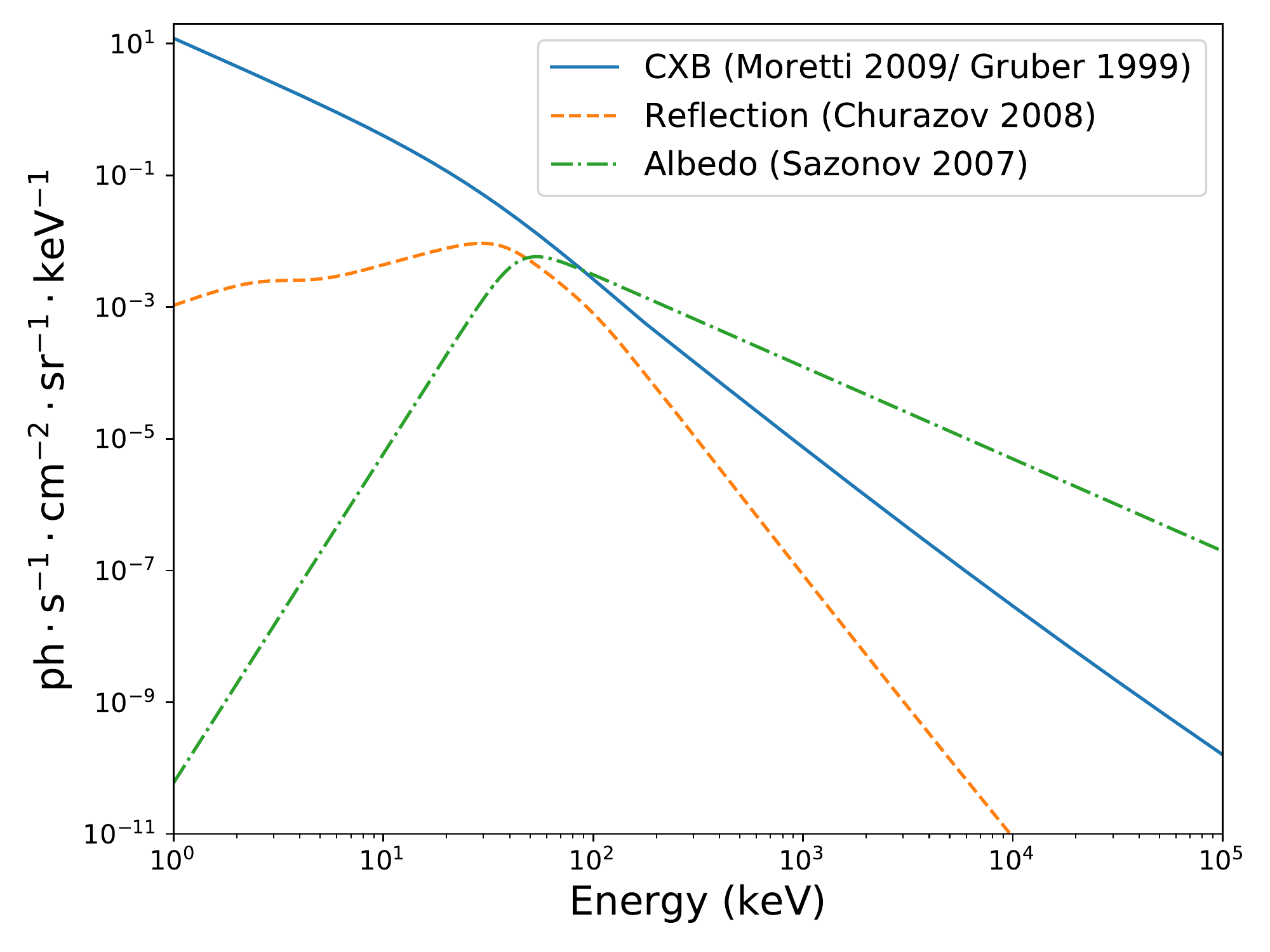}
\caption{The photon spectra of the CXB (Blue), Reflection (Orange) and Albedo (Green) emissions used in our simulations. The Albedo emission spectrum corresponds to the maximum emission observed at latitude of 30$^\circ$N and longitude of 287$^\circ$E.}\label{fig:inp}
\end{figure}

The photon spectrum of the CXB, Reflection and Albedo components are shown in Figure~\ref{fig:inp}.

\subsubsection{CXB}\label{sec:3.1.1}
To describe the photon spectrum of the CXB, we combine the spectral model of ~\cite{Moretti2009} (Equation~\ref{eqn:cxb1}) for photon energies below 163 keV and~\cite{Gruber1999}(Equation~\ref{eqn:cxb2}) for energies above 163 keV. The combined photon spectrum (in units of $\mathrm{ph \cdot cm^{-2} \cdot s^{-1} \cdot sr^{-1} \cdot keV^{-1}}$) is given as:

\begin{subequations}\label{eqn:cxb}
\begin{numcases}{\frac{dN_{CXB}}{dE}=}
   \frac{0.109}{\left(\frac{E}{29.0}\right)^{1.40}+\left(\frac{E}{29.0}\right)^{2.88}} \hspace{15pt} E < 163\ \mathrm{keV} \label{eqn:cxb1}\\[10pt]
  0.0259 \left(\frac{E}{60}\right)^{-6.5} +0.504 \left(\frac{E}{60}\right)^{-2.58} \label{eqn:cxb2} \\
  +0.0288 \left(\frac{E}{60}\right)^{-2.05}\hspace{15pt} E \geq 163\ \mathrm{keV} \nonumber
\end{numcases}
\end{subequations}

\subsubsection{Reflection}\label{sec:3.1.2}
\cite{ Churazov2008} provide a convenient fitting formula to compute the reflection of the CXB photons on the Earth's atmosphere. The outgoing photons spectrum (in units of $\mathrm{ph \cdot cm^{-2} \cdot s^{-1} \cdot sr^{-1} \cdot keV^{-1}}$) can be written as:

\begin{subequations}\label{eqn:ref}
\begin{align}
  \frac{dN_{ref}}{dE} = A(E) \times \frac{dN_{CXB}}{dE} \label{eqn:ref2}
\end{align}
with 
\begin{equation}
  A(E) = \frac{1.22}{(\frac{E}{28.5})^{-2.54} + (\frac{E}{51.3})^{1.57} - 0.37} \times \frac{2.93 + (\frac{E}{3.08})^4}{1 + (\frac{E}{3.08})^4} \times \frac{0.123 + (\frac{E}{91.83})^{3.44}}{1 + (\frac{E}{91.83})^{3.44}} \label{eqn:ref1}\\[5pt]    
\end{equation}
\end{subequations}
where A(E) is the energy dependent reflection factor of the CXB photons.

\subsubsection{Albedo}\label{sec:3.1.3}
\cite{Sazonov2007} provide a formula to calculate the hard X-ray surface brightness of the atmosphere (Albedo) when observed from a spacecraft, provided that its altitude is higher than $\sim 100$ km. The Albedo intensity depends on the satellite position relative to the Earth. The Albedo spectrum (in units of $\mathrm{ph \cdot cm^{-2} \cdot s^{-1} \cdot sr^{-1} \cdot keV^{-1}}$) is given as:
\begin{subequations}\label{eqn:alb}
\begin{align}
 \frac{dN_{alb}}{dE} &= \frac{C}{(\frac{E}{44})^{-5} + (\frac{E}{44})^{1.4}} \label{eqn:alb1}\\[3pt]
C = \frac{3\mu(1 + \mu)}{5\pi} &\frac{1.47 \times 0.0178 \times [(\varphi/2.8)^{0.4} + (\varphi/2.8)^{1.5}]^{-1}}{\sqrt{1 + [\frac{R_{cut}}{1.3\varphi^{0.25}(1 + 2.5\varphi^{0.4})}]^2}} \label{eqn:alb2}\\[3pt]
\intertext{where C has units $\mathrm{ph \cdot cm^{-2} \cdot s^{-1} \cdot sr^{-1}}$ and $\mu = \cos\varTheta$ with,}
 R_{cut} (\mathrm{in\ GV}) &= 59.4 \times \frac{\cos^4\lambda_m}{[1 + (1 + \cos^3\lambda_m \sin\varTheta \sin \xi)^{1/2}]^2} \label{eqn:alb3}
\end{align}
\end{subequations}
Here, $\varphi$ is the solar modulation potential~\citep{Usoskin2005}, $R_{cut}$ the geomagnetic cut-off rigidity, $\lambda_m$ the magnetic latitude and $\varTheta$ is the zenith angle (geographical frame) and  $\xi$ an azimuthal angle, with $\xi= \pi/2$ indicating the North - West direction. Following the  simplification suggested by~\cite{Sazonov2007}, we assume $\xi = 0$. It is to be noted that the measurements of Albedo from SWIFT/BAT data agree well in shape with  the model of~\cite{Sazonov2007}, but differ by a factor 2-3 in terms of normalisation~\citep{Ajello2008}.

\subsubsection{SAA charged particles}\label{sec:3.1.4}

The models of~\cite{Perry2008} and~\cite{Ginet2007} are used to compute the electron and proton spectra respectively. These spectra depend on the geographical position of ECLAIRs in the SAA. To generate these spectra as a function of the \textit{SVOM} orbit, the Systems Tool Kit (STK)~\footnote{\url{http://www.agi.com/home}} software is used.

\subsection{The ECLAIRs mass model}\label{sec:3.2}
\begin{figure}
\centering
\includegraphics[width=0.25\textwidth]{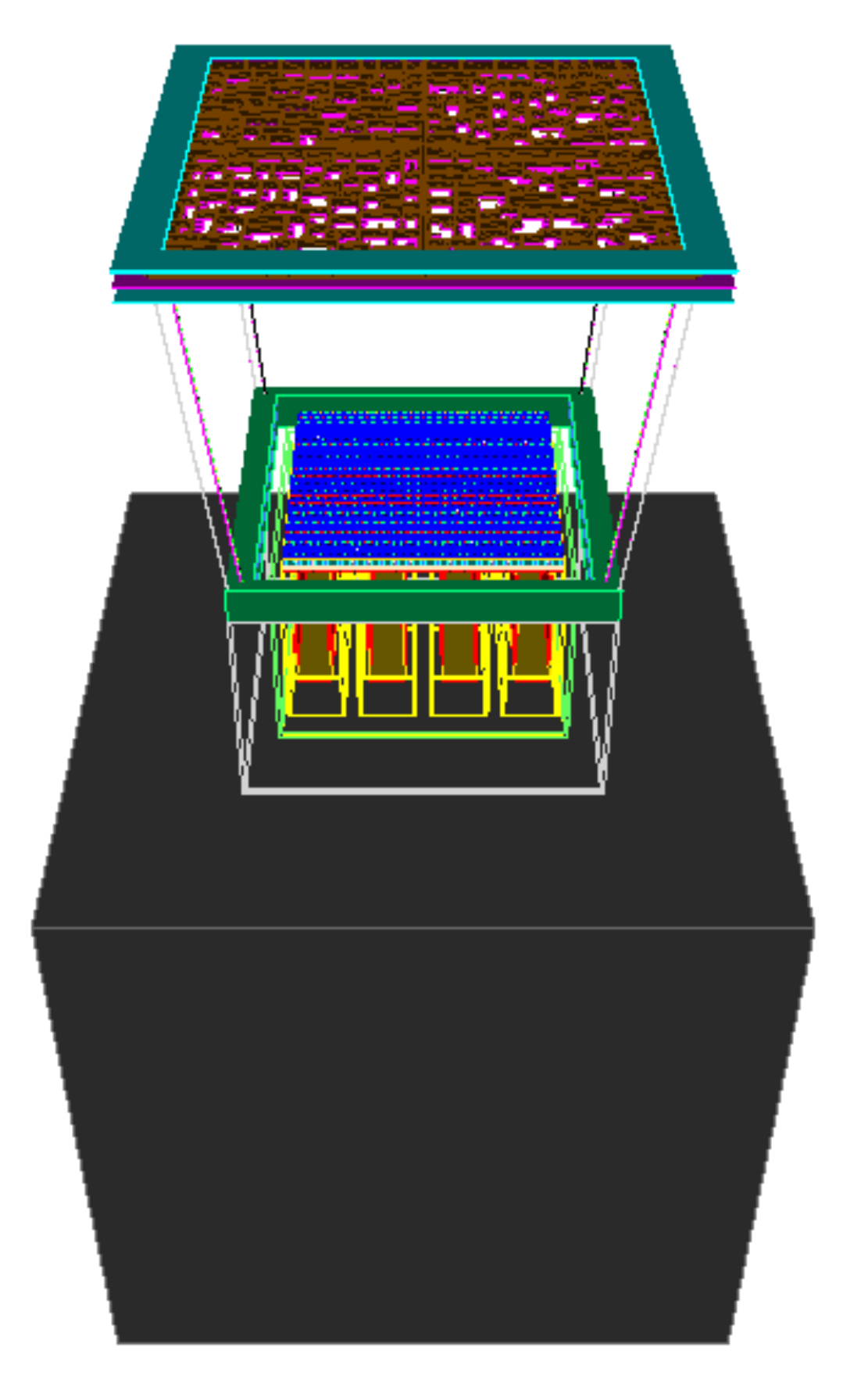}
\caption{Graphical rendering of the ECLAIRs mass model. The mask (Pink), the detector plane (Blue) and the boxes containing electronic readout (Yellow) are shown in the image. The shield surface is not rendered to make internal components visible. The satellite platform (Dark Grey) is roughly approximated by a cube.}\label{fig:eclmm}
\end{figure}

Figure~\ref{fig:eclmm} shows the ECLAIRs mass model used in our simulations. The mass of ECLAIRs is $\sim 87$ kg and of \textit{SVOM} is $\sim 930$ kg. The satellite platform is approximated as a 1 m$^3$ Aluminium cube with a mass of 330 kg. The entire model is contained in a cube of dimensions 2 m $\times$ 2 m $\times$ 2 m. The coded mask (with open fraction of $\sim 0.41$) is a sandwich of Ti--Ta--Ti (Titanium and Tantalum) layers. The coded mask is reinforced with a cross-like structure to ensure its mechanical rigidity. The multi-layer shield (surrounding the mask and detector system) is assembled with Carbon Fiber Reinforced Polymer (CFRP)--Aluminium Honeycomb--CFRP--Lead--Copper layers. The multi-layer insulation (MLI) is composed of Kapton and a thin Aluminium layers. One MLI is placed on top of the mask and two are placed on top of the detector plane. The detector plane consists of 6400 ($80 \times 80$) CdTe pixels each having a size of 4 mm $\times$ 4 mm $\times$ 1 mm. More details are given in the reports~\citep{Sizun2011, Sizun2015}.

\subsection{Building database for ECLAIRs}\label{sec:3.3}
We build separate databases for each of the background component discussed in Section~\ref{sec:3.1}. The simulations are carried out using GEANT4 version 4.10.03. The spatial distribution of primaries is set to be isotropic for all components. The parameter C in Equation~\ref{eqn:alb1} is set to a constant to ignore Albedo emission the angular dependence. The photon energy distributions of the CXB, Reflection and Albedo are given by Equation~\ref{eqn:cxb},~\ref{eqn:ref} and~\ref{eqn:alb1} respectively. We use an ad-hoc average spectral shape to build the database for electrons and protons.

For each component, we store the parameters about the primaries that interact with the detector plane as described in Section~\ref{sec:2.1.2}. We generate a large number of primaries so that the databases contain about $5\times10^7$,  $5\times10^6$, $1\times10^7$ , $1\times10^7$ and $5\times10^5$ detected primaries for the CXB, the Reflection and Albedo, the electrons and the protons respectively.

\subsection{ECLAIRs static background and comparison of PIRA with direct GEANT4 simulations}\label{sec:3.4}

The ECLAIRs background obtained using PIRA in the static case is shown in Figure~\ref{fig:allspec} and the Albedo and SAA charged particles spectra are compared to the related direct GEANT4 simulations.

\subsubsection{ECLAIRs static background for different Earth zenith angles}\label{sec:3.4.1}
To obtain the CXB, Reflection and Albedo background, we carry out simulations using PIRA for several Earth zenith angles $\theta_E$ (with respect to the ECLAIRs frame). The azimuth angle $\phi_E$ is set to be a constant ($\phi_E = 0^\circ$). The satellite altitude ($H$) is fixed to 623 km, the Earth radius ($R_E$) to 6370 km and the atmosphere height ($H_A$) to 100 km. The projected radius of the visible atmospheric spherical cap is $\sim$2400 km. For Albedo, we fix the satellite position where the maximum emission is observed (e.g. latitude $lat_{sat}=30^\circ$N and longitude $lon_{sat}=287^\circ$E)~\footnote{The geomagnetic South pole is located at geographic latitude $80.367^\circ$N and longitude $72.624^\circ$W (reference year 2015).}. This position, together with solar modulation potential ($\varphi$) of 0.25 GV corresponds to the most unfavourable case (maximum emission) and eases qualitative comparisons with the previous work of~\cite{Zhao2012}.

\begin{figure}
\centering
\includegraphics[width=\textwidth]{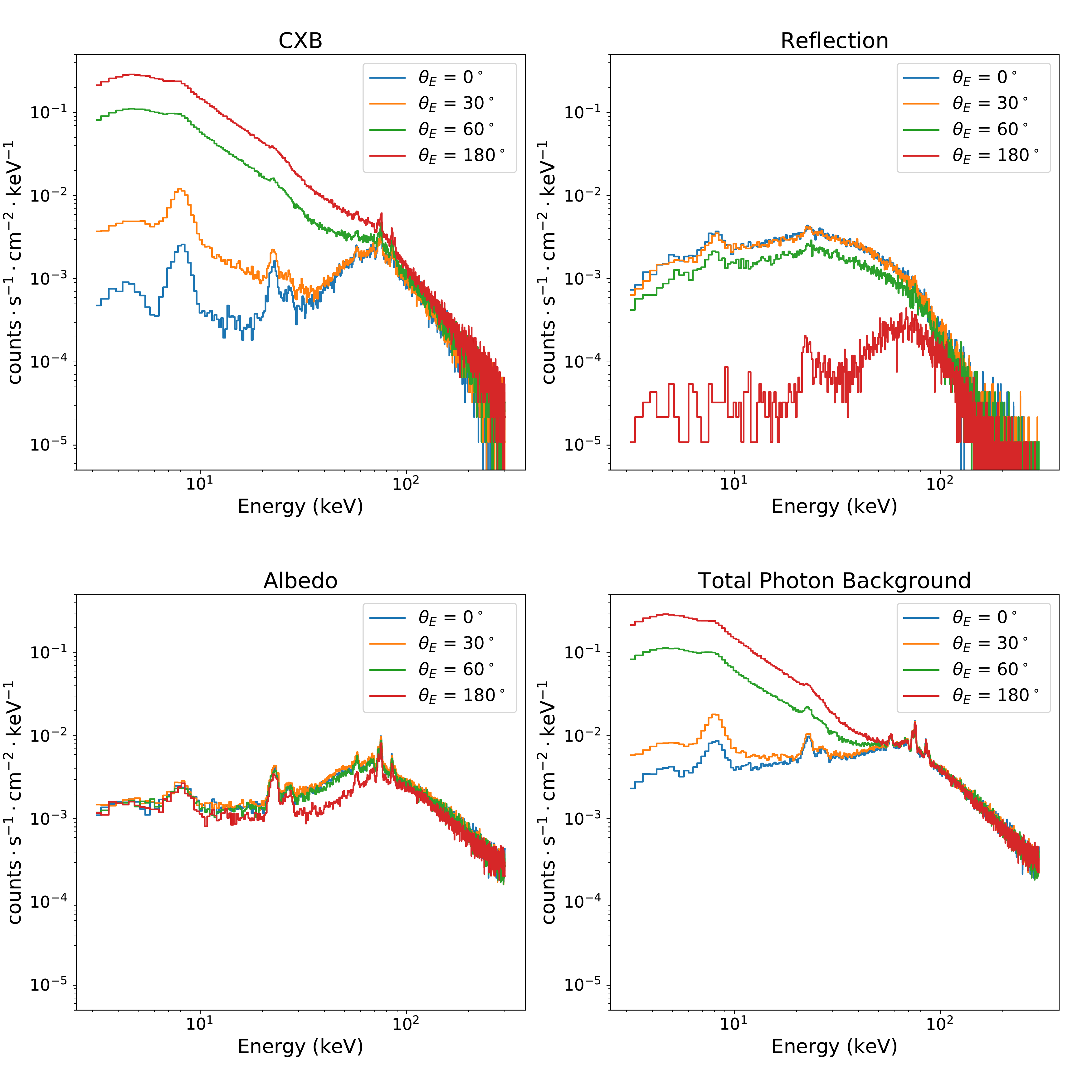}
\caption{ECLAIRs spectra of all the photon background components for different Earth zenith angles : CXB, Reflection and Albedo on the top left,  top right and bottom left respectively. The Albedo is estimated at the position where the maximum emission is observed ($30^\circ$N and $287^\circ$E). The total spectrum (sum) of all photons components is shown in the bottom right panel.}\label{fig:allspec}
\end{figure}

The CXB, Reflection and Albedo spectra, and the total spectrum (sum of these components) for different Earth zenith angles obtained using PIRA are shown in Figure~\ref{fig:allspec}. The emission lines induced by the fluorescence of instrument body elements (e.g. materials in the mask, shield, the detectors etc) are clearly seen in all the background spectra. These lines are useful for the in-flight energy calibration of the detectors. Altogether, we have a good qualitative agreement between the PIRA spectra and previous GEANT4 simulations of~\cite{Zhao2012}.

\subsubsection{Case of Albedo}\label{sec:3.4.2}
\begin{figure}
\centering
\includegraphics[width=.75\textwidth]{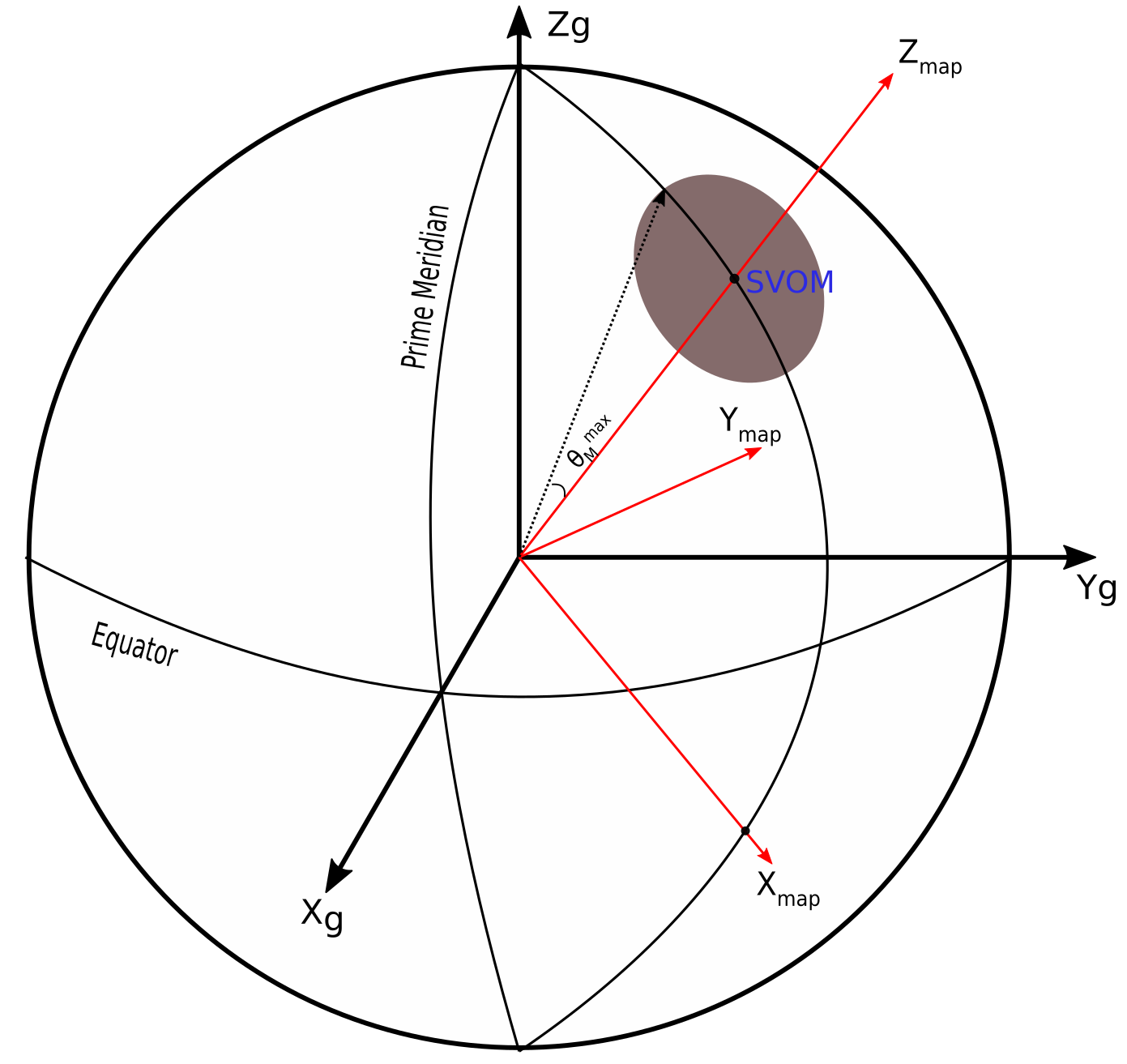}
\caption{Schematic diagram showing the relation between the Geographical frame (latitude/longitude) and the ``MAP" frame (Figure~\ref{fig:albmap}) for a given satellite position. The subscript ``g" denotes the Geographical frame while the subscript ``map" denotes the ``MAP" frame.}\label{fig:MAP2Geo}
\end{figure}

The Albedo emission depends on the geographic position (latitude and longitude) of the satellite (Equation~\ref{eqn:alb}). To compute the emission intensity $I(\theta,\phi)$ for a given satellite position, we  transform the ``MAP" coordinates $\theta_m$ and $\phi_m$ into latitude and longitude coordinates. The orientation of the ``MAP'' frame depends on the satellite position (Figure~\ref{fig:MAP2Geo}). We assume that the X-axis of the ``MAP" frame is parallel to the local geographical longitude of the satellite. Then, the transformation can be performed with the rotation matrix (defined by ZYZ rotation rule,~\cite{Henderson1977}) with the proper Euler angles ($lon_{sat}$, $90^\circ - lat_{sat}$, $0^\circ$). The geographical latitude and longitude are further converted to the geomagnetic latitude and longitude (Equation~\ref{eqn:alb3}).

\begin{figure}
    \centering
    \includegraphics[width=.9\textwidth]{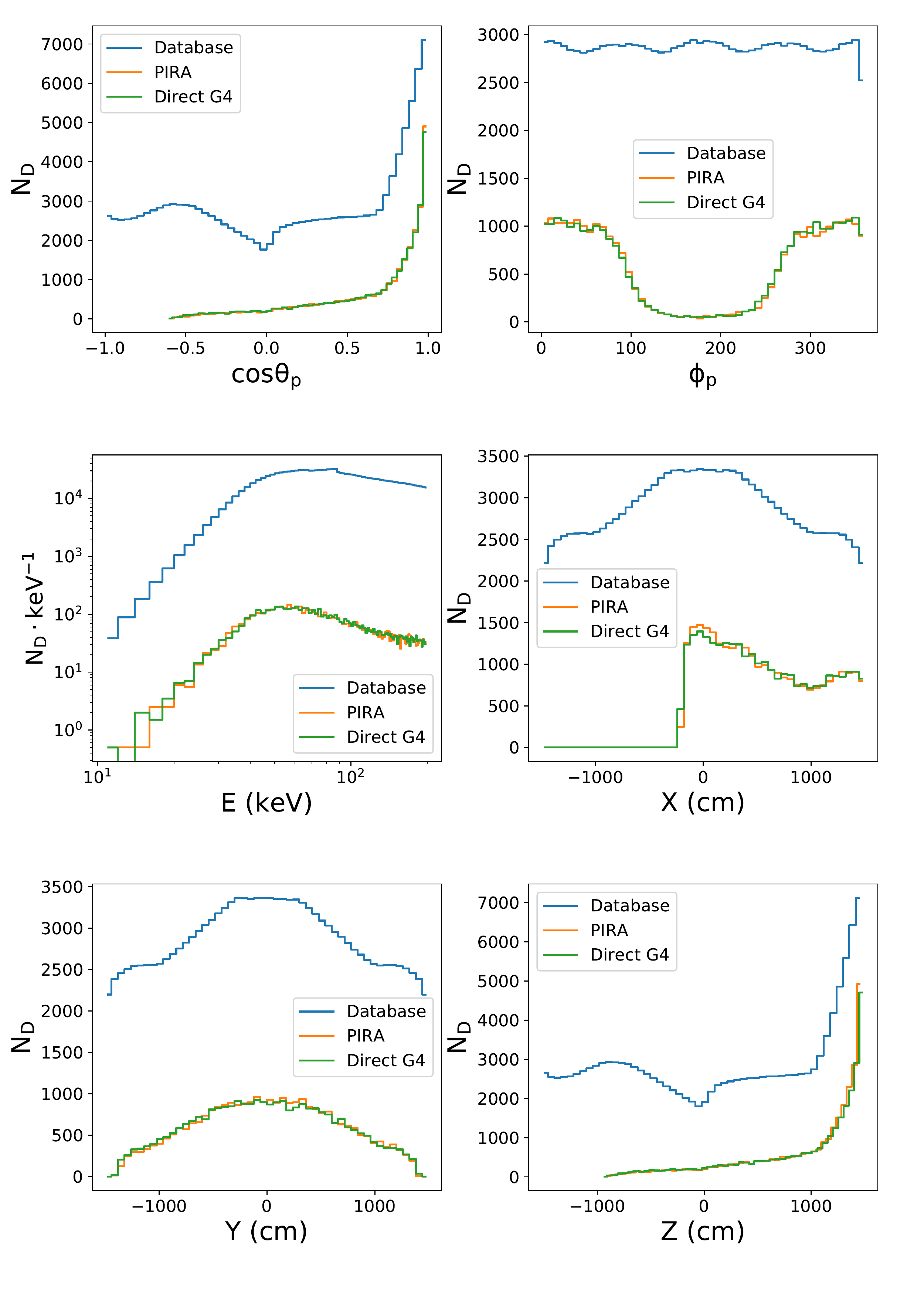}
    \caption{Distribution of detected Albedo photons ($N_D$) using PIRA (Orange) and the related direct GEANT4 simulations (Green), for $60^\circ$  Earth zenith angle. Distribution of incoming photon's zenith angle (Top-left) and azimuth angle (Top-right). Distribution of incoming photon's energy (Middle left). Initial photons position distribution (Middle right and bottom panels). In all the plots, ``Database" (Blue) is the distribution of all the detected photons stored in the Albedo database (without any selections). It is arbitrarily scaled down to ease a comparison.}\label{fig:albinp}
\end{figure}

\begin{figure}
    \centering
    \includegraphics[width=\textwidth]{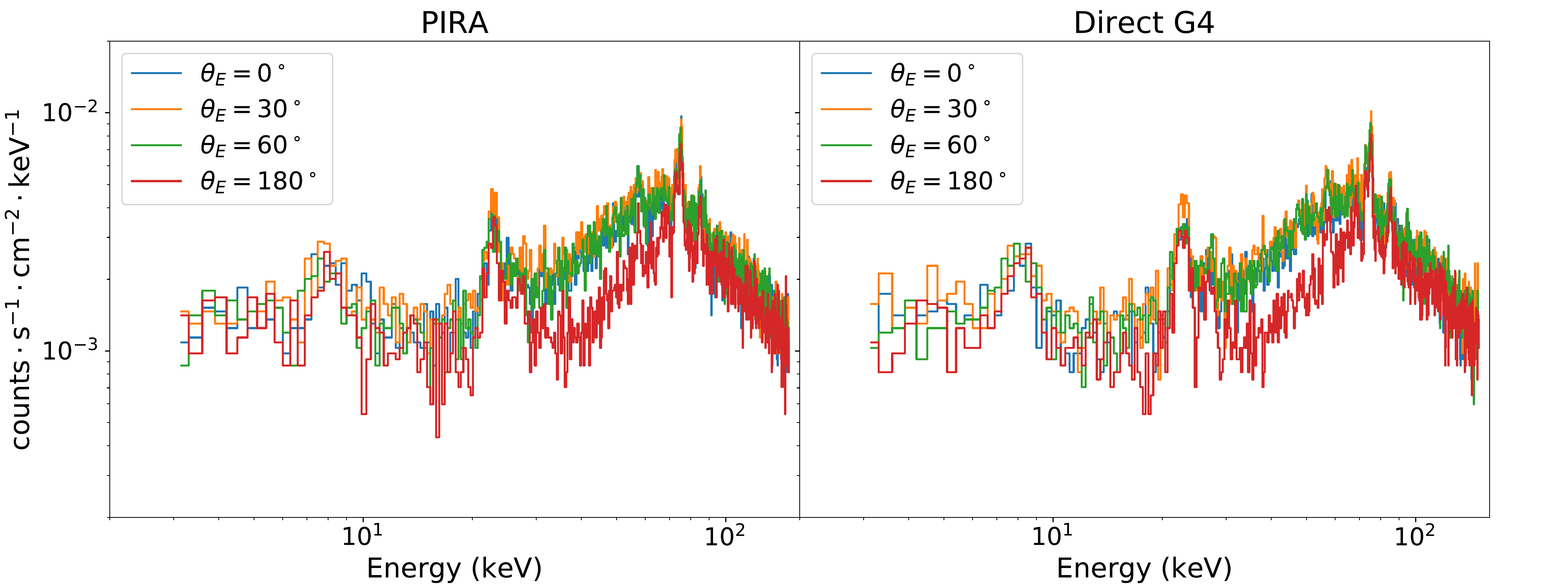}
    \caption{Albedo background spectra for several Earth zenith angles. (Left) the output of PIRA, (Right) the output of a direct GEANT4 simulation. In both cases the Earth azimuth angle ($\phi_E$) is equal to  $0^\circ$.}\label{fig:9}
\end{figure}

\begin{figure}
    \centering
    \includegraphics[width=\textwidth]{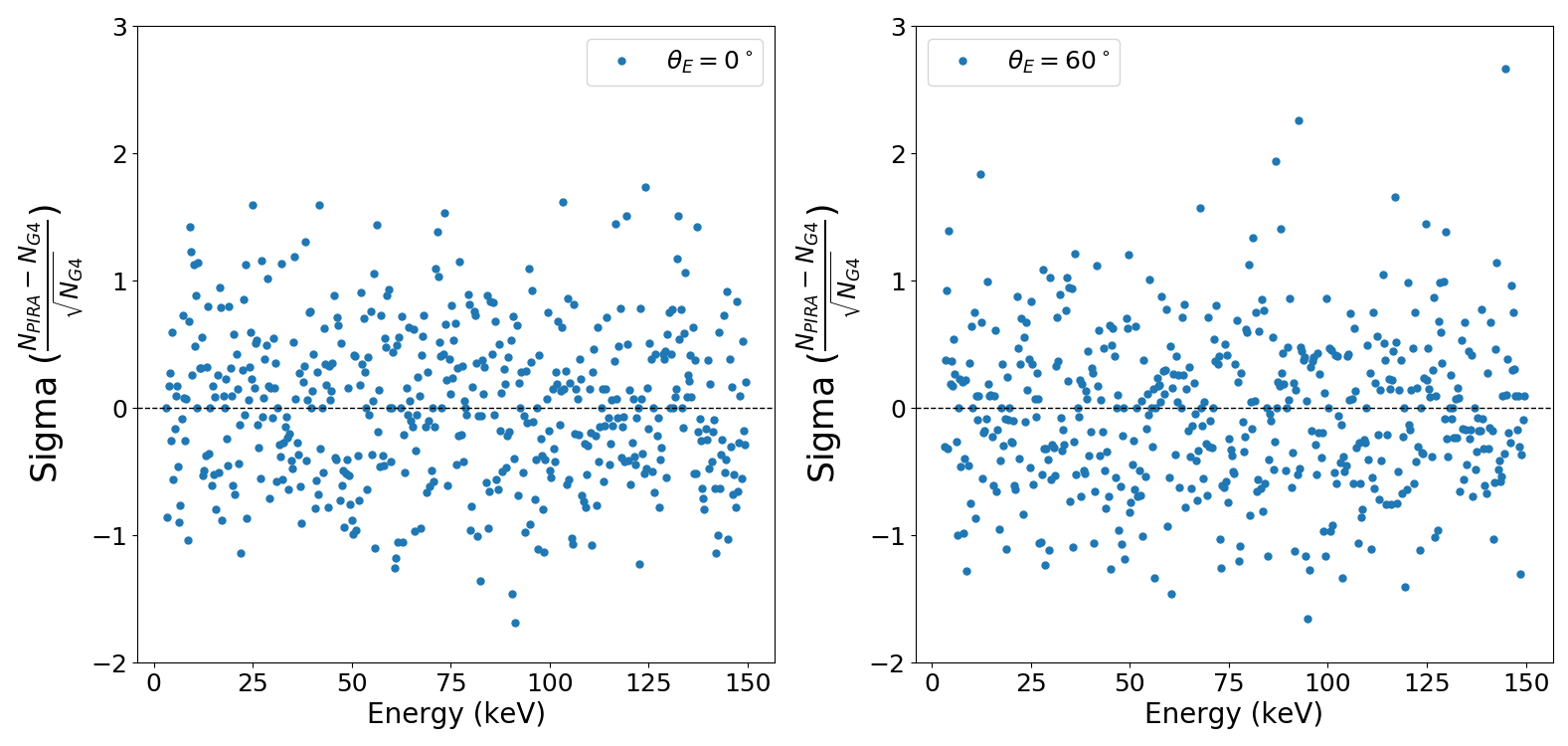}
    \caption{Residuals between the albedo spectrum obtained with PIRA and direct GEANT4 (Figure~\ref{fig:9}) for two Earth zenith angles ($0^\circ$ and $60^\circ$). Here, $\mathrm{N_{PIRA}}$ and $\mathrm{N_{G4}}$ are the number of counts per energy bin respectively for the PIRA and the direct GEANT4 albedo spectrum.}\label{fig:residuals}
\end{figure}

For $\theta_E  = 60^\circ$ and $\phi_E= 0^\circ$ we compare the spatial distribution of selected photons on the source sphere (Figure~\ref{fig:albinp}). This includes the distribution of initial directions ($\theta_p, \phi_p$), initial positions ($x_p$, $y_p$, $z_p$) on the source sphere and the spectral distribution of  the detected photons. In addition, we compare the PIRA spectrum with the related direct GEANT4 simulations for several Earth zenith angles. Spectra obtained with both methods show relatively good  agreement (Figure~\ref{fig:9} and~\ref{fig:residuals}).


\subsubsection{Case of SAA charged particles}\label{sec:3.4.3}
\begin{figure}
 \centering
 \includegraphics[width=.8\textwidth]{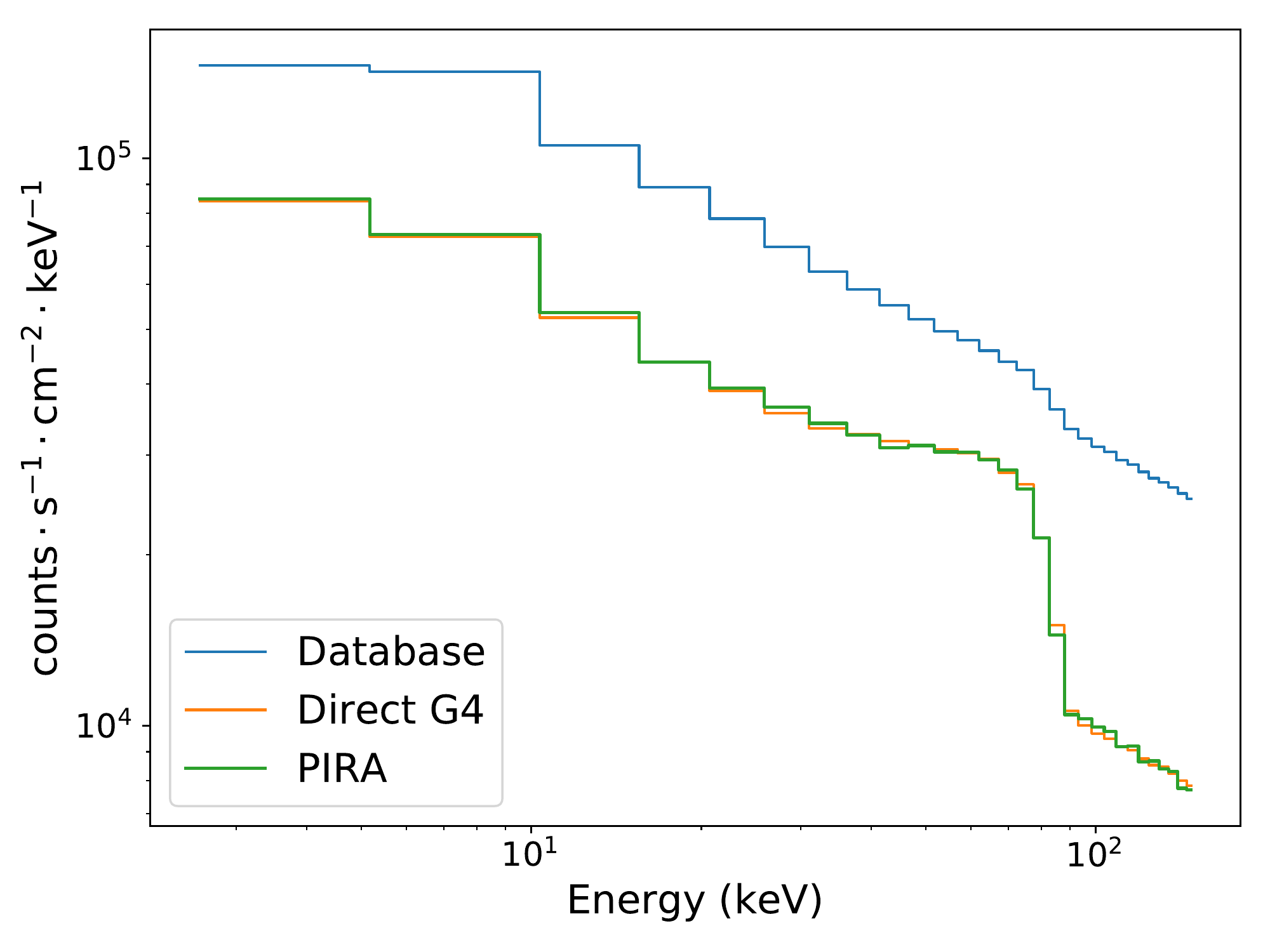}
 \caption{Spectra obtained using PIRA (Green) and direct GEANT4 simulations (Orange) for a given input electron spectrum. The ``Database" (Blue) is the spectrum of all the events stored in the electron database (arbitrarily scaled down to ease a comparison}\label{fig:cpver}
\end{figure}

In order to verify that PIRA is able to model different spectral forms, notably in the SAA, we generate several incoming spectra of electrons and protons. For the same input spectrum (with fixed intensity and shape), we compare direct GEANT4 simulations and PIRA results. In all cases, the spectra obtained by both methods are in good agreement. This comparison is shown in Figure~\ref{fig:cpver} for an electron spectrum.

\subsection{ECLAIRs dynamic background}\label{sec:3.5}
To  perform  dynamic simulations of the background events for ECLAIRs, we need to know the satellite position as a function of time. We rely on the predicted \textit{SVOM} orbital parameter computation carried out by~\cite{Jaubert2017}. These parameters include the geographic latitude and longitude, the satellite altitude and Earth directions and the roll angle ($\psi_E$,  defined as the azimuthal angle of the vector pointing towards the local north direction).

\subsubsection{Photon background for ECLAIRs}\label{sec:3.5.1}
\begin{figure}[ht!]
 \centering
 \includegraphics[width=\textwidth]{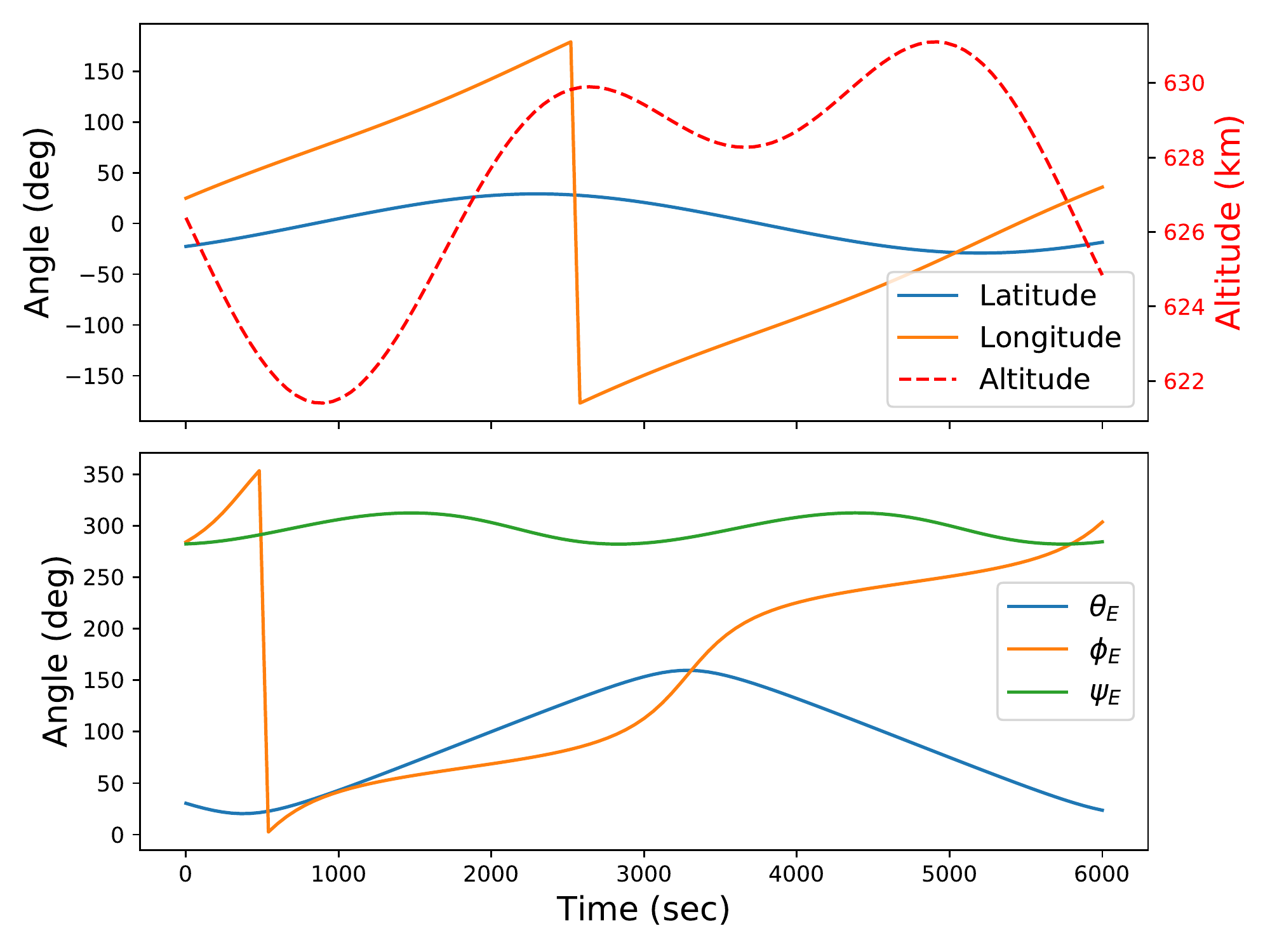}
 \caption{Evolution of the orbital parameters as a function of time. Top-panel: latitude (Blue), longitude (Orange) and altitude (Red dashed line). Bottom-panel: the Earth directions $\theta_E$ (Blue), $\phi_E$(Orange) and the roll angle $\psi_E$ (Green). For this orbit, the ECLAIRs pointing direction coordinates are RA = 158.06$^\circ$, Dec = 31.64$^\circ$, Roll = 94.86$^\circ$}\label{fig:orbit}
\end{figure}

\begin{figure}
    \centering
    \includegraphics[width=\textwidth]{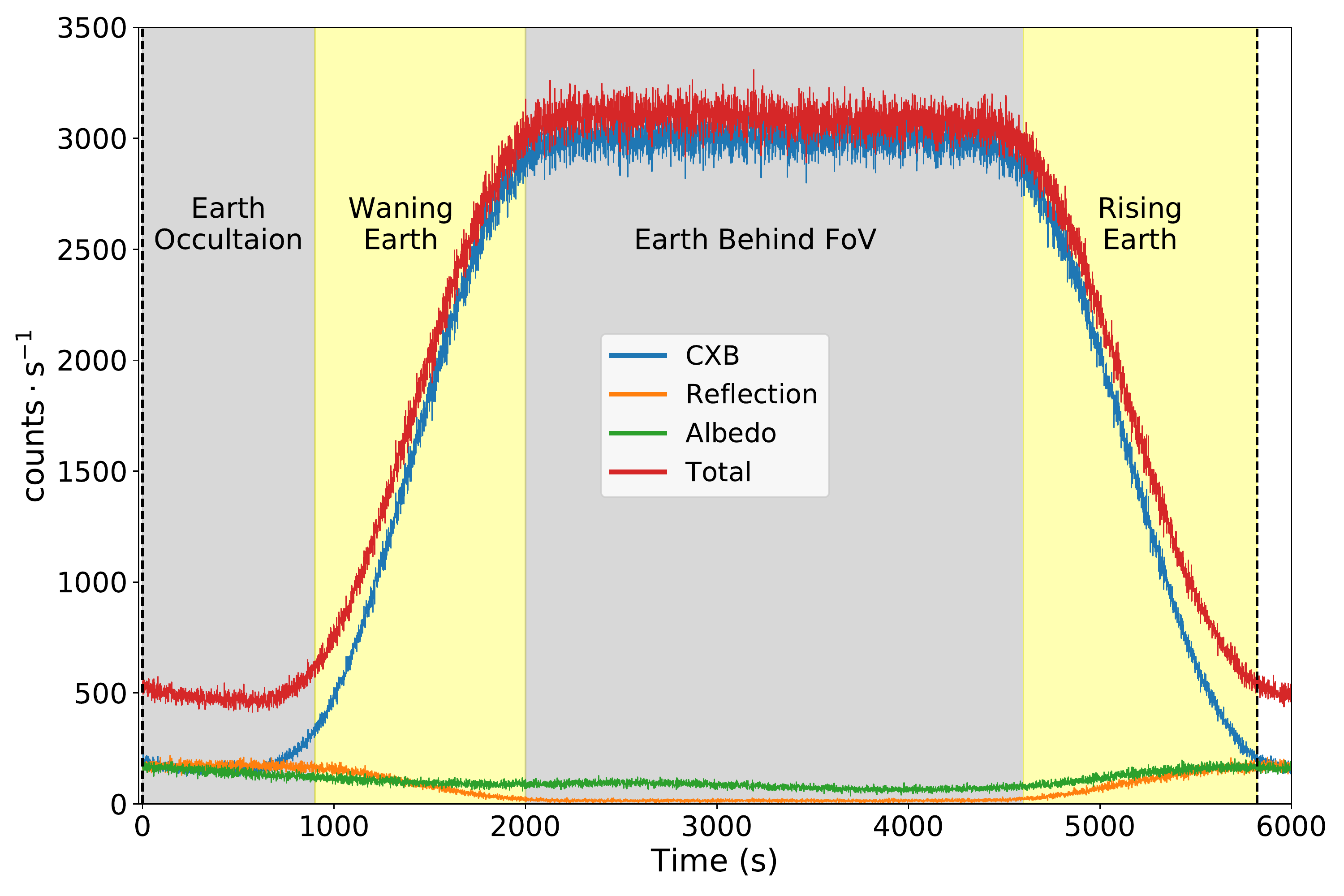}
    \caption{Typical modulations of the photon background: CXB (Blue), Reflection (Orange) and Albedo (Green). The orbit duration ($\sim 98$ minutes) is delimited by the vertical dotted lines. The corresponding orbital parameter variations are shown in Figure~\ref{fig:orbit}. The lightcurve is built using single events only (see Section~\ref{sec:2.1.2}) in the 4--150 keV energy range.}\label{fig:13}
\end{figure}

Figure~\ref{fig:orbit} shows the orbital parameter variation during a typical \textit{SVOM} orbit ($\sim$98 min). During such an orbit the ECLAIRs pointing direction (Right Ascension, Declination and Roll angle) is kept fixed. Figure~\ref{fig:13} displays the evolution of the instrument background and of all the  photon components during the same orbit.

The CXB is the dominant background component with about $\mathrm{\sim 3000\ counts \cdot s^{-1}}$ when ECLAIRs is pointing away from the Earth. For this orbit, the CXB is heavily modulated by the Earth occultation of the FoV. When the Earth completely obscures the field of view, the CXB, Reflection and Albedo intensities are comparable. Since the Albedo intensity depends on the satellite geographic position, its variation is not completely correlated with the presence of the Earth in the FoV.

\begin{figure}
    \centering
    \includegraphics[width=.8\textwidth]{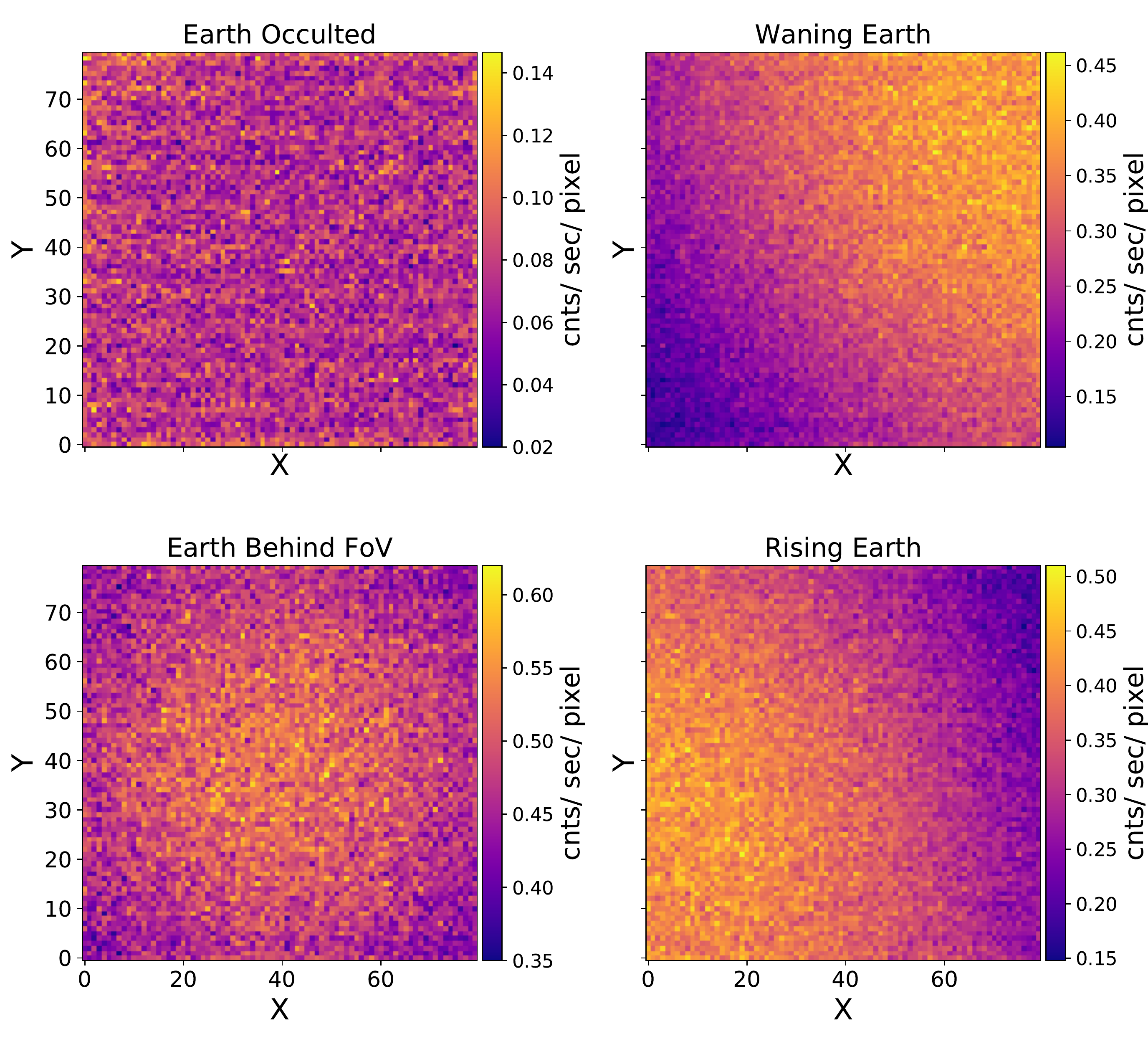}
    \caption{Detector plane images or shadowgrams (single events only) for periods marked with the shaded area in Figure~\ref{fig:13}. The structures in the Earth  image when the Earth obscures the field-of-view (top left) are formed by high energy photons coming from the bottom of the instrument as a result of the simplified model of the satellite platform.}\label{fig:14}
\end{figure}

The background variations also have a significant impact on the spatial distribution of photons on the detector plane. Figure~\ref{fig:14} shows the detector plane images (shadowgrams) at different epochs along the orbit (time intervals are indicated in Figure~\ref{fig:13}).

\subsubsection{SAA background for ECLAIRs}\label{sec:3.5.2}

\begin{figure}[ht!]
    \centering
    \includegraphics[width=.7\textwidth]{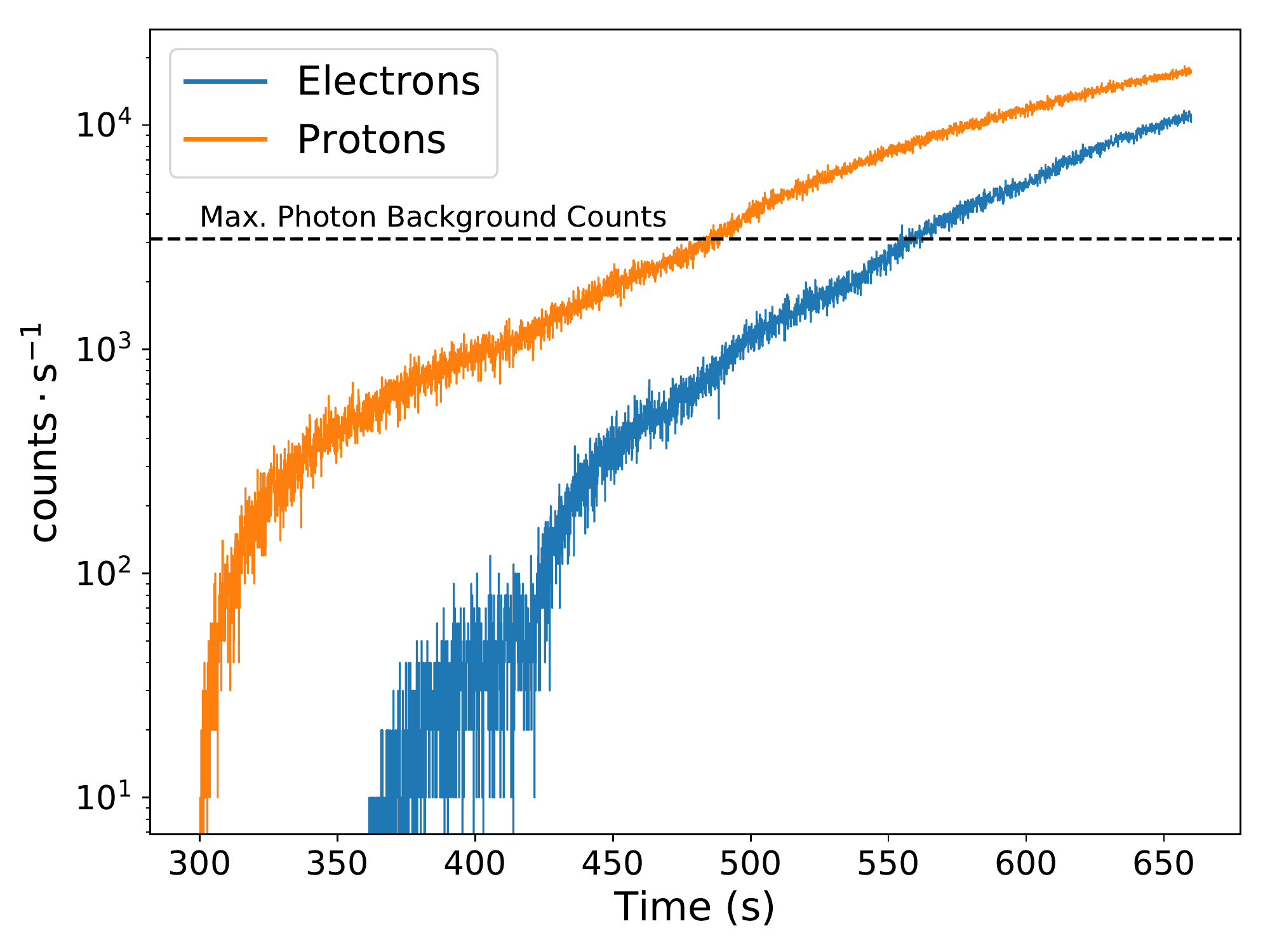}
    \caption{Typical variation of the background (single events only) due to the electrons (Blue) and protons (Orange) when ECLAIRs enters the SAA. For comparison, the maximum photon background count rate (sum of the CXB, Reflection and Albedo components) is shown as the dotted line.}\label{fig:15}
\end{figure}

For safety reasons, ECLAIRs will not perform observations while passing through the deep SAA. However, we are interested in the count rate increase and decrease at the entrance and the exit of the SAA in order to put constraints on the operations ({e.g. to test the constraints on the onboard trigger operations}). Figure~\ref{fig:15} shows the increase of both detected electrons and protons when the satellite enters the SAA. It can be seen that the count rate reaches values of $\mathrm{\sim10^4\ counts \cdot s^{-1}}$ within 5 minutes, both for electron and proton components. However, it should be noted that the count rate and its rate of change heavily depends on the model used to determine the input spectra of the charged particles.

\subsection{Comparison of computation time}\label{sec:3.6}
We compare the time to perform one minute and one orbit ($\sim$98 minutes) duration simulations of the CXB and Albedo, in the static case (see Table~\ref{tab:4}). All the simulations are performed on a single core of a 2.2 GHz processor computer. The CXB and Albedo emission are set to their maximum emission. From Table~\ref{tab:4}, we clearly see that the PIRA method speeds up the computation process by a factor of $\sim 10^3 - 10^4$ as compared to a direct GEANT4 simulation. 

It should be noted that the computation of databases can take large amount of time. However, once it has been created, simulations for any considered orbital configuration are performed very quickly using PIRA.

\begin{table*}
\caption{Comparison of the time taken by PIRA and a direct GEANT4 simulation to compute the background in the static case.}\label{tab:4}
\begin{tabularx}{\textwidth}{|Y|c|c|c|}\hline
\multirow{2}*{Duration}&\multirow{2}*{Background Component}&\multicolumn{2}{c|}{Computation Time (in minutes)}\\
\cline{3-4}
& & Direct GEANT4 & PIRA \\ \hline
\multirow{2}*{1 min} & CXB & 951 & 0.58 \\ \cline{2-4}
& Albedo & 925 & 0.38\\\hline
\multirow{2}*{98 min} & CXB & 93150.88 & 5.9 (4.42)$^\ast$ \\ \cline{2-4}
& Albedo & 90625.75 & 16.28 (6.68)$^\ast$\\
\hline
\end{tabularx}\\
$^\ast$ The value in the parenthesis corresponds to the computation time in the dynamic case i.e, when the orbital parameters are evolving with time.
\end{table*}

\section{Discussions and conclusion}\label{sec:4}
We developed the PIRA algorithms to simulate the ECLAIRs background events with the aim of avoiding the computation burden of performing a new GEANT4 simulation every time the orbital configuration of \textit{SVOM} changes. It is a general method which can simulate the background for varying attitude parameters i.e. a complete dynamic simulation.

Concerning the creation of the databases of primaries, it is also possible to create a unique database for all the photon components (i.e. for CXB, Reflection and Albedo). However, the resulting algorithms are more complex and require a combination of spatial mapping (see Section~\ref{sec:2.3.1}) and spectral mapping (see Section~\ref{sec:2.3.2}) in all cases to select photons. To keep the algorithms simple, we decide to create a separate database for each component.

A special care must be taken while using the primaries in the database. In order to keep the statistical properties of the output spectra, we should not pick the same interacting primary more than once for a given PIRA simulation. This condition restricts the duration of the PIRA simulation. For example, the $5\times10^7$ interacting primaries stored in the ECLAIRs CXB database can be used to simulate the CXB events for a maximum duration of 216 minutes (approximately two \textit{SVOM} orbits). If a larger duration is required, we just need to increase the number of detected primaries stored in the database.

In the case of charged particles, the generation of the database is more time consuming compared to the photon database due to the many secondary interactions. For the SAA, we are only interested in the count rate variation (see Section~\ref{sec:3.5.2}) hence we relax the condition of only picking up the non-repeating primaries from the database.

The dynamic background simulations of ECLAIRs (carried out using PIRA) are used to generate simulated raw data (with addition of GRBs and quiescent X-ray sources). With PIRA, the generation of such raw data is possible in a very short time. It enables simulation of realistic and complex observing test cases prior to the launch. This will be helpful in the scientific preparation of the mission (e.g testing onboard and offline trigger software, ground data reduction software). In addition, we also plan to include the cosmic-ray and atmospheric neutron contribution to the background using PIRA. A preliminary calculation of the cosmic-rays background assuming protons, alpha particles, Carbon, Oxygen and Iron composition has been done for ECLAIRs. In the 4 - 250 keV energy range, the cosmic-rays particles contribute to the ECLAIRs background at the rate of $\sim \mathrm{200\ counts\cdot s^{-1}}$, to be compared to the $\sim \mathrm{3500\ counts\cdot s^{-1}}$ due to the CXB~\citep{Sauvageon2009}.

Although the PIRA algorithms are developed for ECLAIRs, they are general and do not have an explicit dependence on the design and characteristics of the instrument. The dependence on the instrument properties is implicitly included in the GEANT4 simulations performed to generate the database. This makes PIRA a general method which can be easily extended to other space instruments like ECLAIRs (in the case of \textit{SVOM}, there are plans to extend the method to the Gamma-ray Monitor (GRM) instrument). 

The gain in computation time (see Section~\ref{sec:3.6}) compared to direct GEANT4 simulations, the general nature of the algorithms and the ability to perform dynamic simulations make PIRA a powerful tool for background simulations of current and future high-energy space instruments.


\bibliographystyle{aasjournal}
\bibliography{article_PIRA.bib}   

%
%

\end{document}